\documentclass[12pt]{iopart}
\usepackage{iopams}
\usepackage{epsf,graphics,graphicx}
\usepackage{float}
\usepackage{subfigure}
\usepackage{xcolor}

\newcommand{\ket}[1]{|#1\rangle}
\newcommand{\bra}[1]{\langle #1|}

\begin{document}
\title{The role of quantum coherence in dimer and trimer excitation energy transfer}
\author{Charlotta Bengtson}
\address{Department of Chemistry - {\AA}ngstr\"om Laboratory, Theoretical Chemistry,
Uppsala University, Box 538, SE-751 21 Uppsala, Sweden}
\ead{charlotta.bengtson@kemi.uu.se}
\author{Erik Sj\"oqvist}
\address{Department of Physics and Astronomy, Uppsala University, Box 516,
SE-751 20 Uppsala, Sweden}
\ead{erik.sjoqvist@physics.uu.se}
\date{\today}
\begin{abstract}
Recent progress in resource theory of quantum coherence has resulted in measures to 
quantify coherence in quantum systems. Especially, the $l_1$-norm and relative entropy 
of coherence have been shown to be proper quantifiers of coherence and have been  
used to investigate coherence properties in different operational tasks. Since long-lasting 
quantum coherence has been experimentally confirmed in a number of photosynthetic 
complexes, it has been debated if and how coherence is connected to the known 
efficiency of population transfer in such systems. In this study, we investigate quantitatively 
the relationship between coherence, as quantified by $l_{1}$-norm and relative entropy of 
coherence, and efficiency, as quantified by fidelity, for population transfer between 
end-sites in a network of two-level quantum systems. In particular, we use the coherence
averaged over the duration of the population transfer in order to carry out a quantitative 
comparision between coherence and fidelity. Our results show that although coherence 
is a necessary requirement for population transfer, there is no unique relation between 
coherence and the efficiency of the transfer process.
\end{abstract}
\submitto{\NJP}
\maketitle
\section{Introduction}
The idea of using and quantifying coherence as a resource was first introduced by 
{\AA}berg \cite{Aberg2006}. Later, Baumgratz {\it et al.} \cite{Baumgratz2014} further developed
the resource theory and framework for quantification of quantum coherence. Their approach 
is directly analogous to entanglement theory; the resource states correspond to entangled 
states while the free (incoherent) states correspond to separable states in entanglement 
theory. Incoherent operations are those which map the set of free states back to itself, 
similar to the preservation of separable states under local operations and classical 
communication (LOCC) in entanglement theory. 

Quantification of quantum coherence, which makes it possible to distinguish different quantum 
states in terms of their ability to function as a coherence resource, is introduced as a set of conditions 
for a functional $C$ mapping quantum states into non-negative real numbers. Any functional 
$C$ satisfying these conditions are classified as a proper coherence measure. Two measures 
are the {\it $l_{1}$-norm of coherence} and the {\it relative entropy of coherence} (REOC) 
\cite{Baumgratz2014}. Following \cite{Aberg2006,Baumgratz2014}, a number of works 
studying different aspects of quantum coherence as a resource have been published \cite{Singh2015_2,Singh2015_1,Girolami2014,Winter2016,Napoli2016,Rana2016,
Streltsov2015,Chitambar2016_1,Chitambar2016_2} 
and several new coherence measures have been proposed \cite{Girolami2014,Napoli2016,Rana2016,Streltsov2015,Shao2015}. The progress 
in the field has been reviewed recently \cite{Streltsov2016}.
Since quantum coherence is a basis-dependent property, measures require a choice 
of a particular basis with respect to which we define the free states and incoherent
operations. In what way we choose our preferred basis depends on what kind of task  
we are studying. 

There are different processes where quantum coherence has been shown to enhance the outcome 
in some aspect, i.e., function as a resource. Examples can be found in quantum thermodynamics \cite{Lostaglio2015a,Kim2015,Korzekwa2016,Misra2016,Chen1_2016} and photocells 
\cite{Creatore2013,Zhang2015}. A mechanism for which the importance 
of quantum coherence has been widely discussed and studied over the last years is excitation
energy transfer (EET) in photosynthetic complexes. Such molecular aggregates typically consist 
of a number of coupled chromophores (light absorbing molecules) in a protein scaffold, where 
a quantum-mechanical excitation, initially located on one chromophore, is transferred to a 
special chromophore in the network. This special chromophore is connected to a reaction 
center, where the excitation is captured and converted to chemical energy. Since photosynthetic 
complexes are known to convert light into chemical energy in a very efficient manner 
\cite{Chain1977}, the discovery of long-lived quantum coherence in the Fenna-Mattews-Olson
complex \cite{Engel2007} started speculations on whether the observed coherence could explain 
the very efficient EET. The initial idea was that quantum coherence would allow the complex to 
perform a quantum computation, analogous to Grover search \cite{Grover1997},
to find the most efficient pathway from the initially excited chromophore to the chromophore in 
contact with the reaction centre. 

Today, it is known that quantum coherence on its own is not enough to produce high quantum 
efficiency in such molecular aggregates. Instead, quantum coherence together with the coupling 
and dynamics of the environment of the system facilitates an efficient EET. Possible mechanisms 
behind such environment-assisted quantum transport have been suggested by including environmental 
effects in different ways \cite{Plenio2008,Mohseni2008,Caruso2009,Chin2010,Wu2010,Irish2014,Rebentrost2009_1}. 
Nevertheless, the existence of quantum coherence during EET in such systems seems to be an 
experimentally established fact 
\cite{Engel2007,Hayes2010,Panitchayangkoon2010,Panitchayangkoon2011}.
It is hence of interest to study the effect of coherence on EET in a quantitative manner. 
A first step is to study a quantum system consisting of a network of sites with no 
environmental interaction and relate the amount of coherence, by making use of certain
coherence measures, to the efficiency of a population transfer between sites. 

Interest in efficient population transfer and its relation to quantum coherence is not limited 
to EET in molecular aggregates; the same idea applies for every quantum system where an 
efficient quantum state transfer between sites in a network is required. Examples of physical 
scenarios that can be described in the same manner are electron transport in a network of 
quantum dots \cite{Greentree2004}, photons in an optical network \cite{Schreiber2010}, 
and spin chains \cite{Bose2007,Kay2010}.

In this study, we investigate quantitatively the role of quantum coherence, as measured by 
$l_{1}$-norm of coherence and REOC, for efficient population 
transfer between the end-sites in a network of either two sites (dimer) or three sites 
(trimer). These systems can be thought of as the primitive units of tunneling between  
sites and interference between different pathways, respectively, which are 
two main physical mechanisms in population transfer. Specifically, we investigate under 
which conditions, in terms of the parameter space of the Hamiltonian of the systems, 
maximal efficiency and maximal coherence are found, and whether these 
maximizing parameter choices coincide. 

The present study can be useful in a general context where quantum coherence as a resource for 
different tasks is investigated, but especially it can reveal features if and how quantum 
coherence can be used to optimize population transfer in a network of sites, as well as 
how the Hamiltonian of such a system would look like. This can in turn provide useful 
information on how to construct quantum networks like artificial photosynthetic complexes 
in the most efficient manner. For instance, in \cite{Hemmig2016,Woller2013} a new technique, 
where chromophores can be placed one-by-one with great precision, has been developed. 
It is hence already possible to create man-made networks of chromophores in the lab.

The outline of the paper is as follows. The system of interacting sites is 
introduced in the next section. Section \ref{sec:Quantifying-coherence-and-efficiency} 
contains a description of the different quantifiers of efficiency and coherence that are 
used in this work. The dimer and trimer cases are analyzed numerically and analytically 
in sections \ref{sec:dimer} and \ref{sec:trimer}, respectively. The long-term behavior of 
the coherence is studied in section \ref{sec:long-term}. The paper ends with the conclusions. 

\section{System}
\label{sec:system}

\begin{figure}
\includegraphics[scale=0.2]{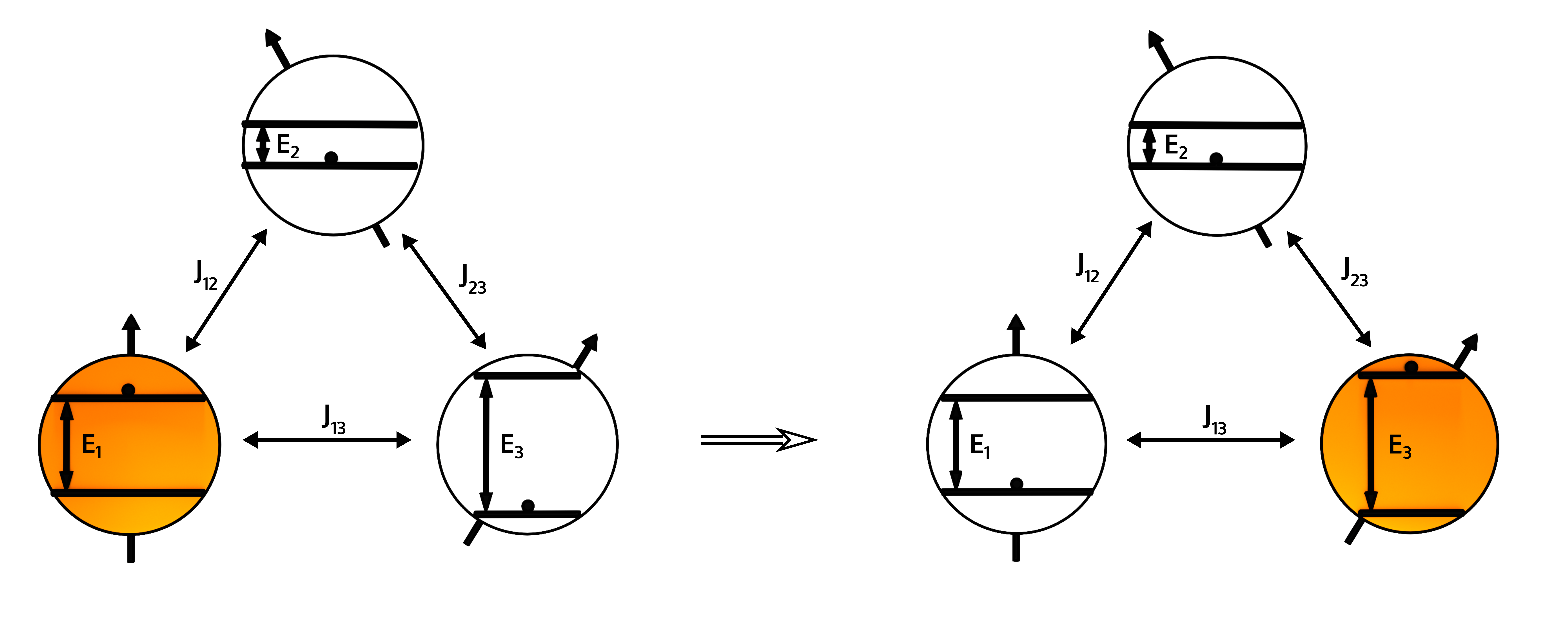}\protect\caption{\label{fig:Trimer-system}
Schematical picture of a three-site (trimer) system. The excitation energy of site $i$ 
is denoted $E_{i}$ and the coupling between sites $i$ and $j$ is denoted $J_{ij}$. The 
arrows symbolize the dipole moments in the case where the sites are occupied by localized 
chromophores. The alignment and distance between these dipole moments determine the 
inter-site couplings.}
\end{figure}

The system considered in this study is a network of $N$ localized sites, which are coupled together 
by their mutual interaction. Each site can be modeled as a two-level quantum system consisting of a 
ground state $\ket{\mathbf{g}}$ and an excited state $\ket{\mathbf{e}}$. A possible realization 
could be a molecular aggregate of chromophores, each localized at a site of the network and 
with $\ket{\mathbf{g}}$ and $\ket{\mathbf{e}}$ being the ground and excited singlet states, 
respectively. Population of $\ket{\mathbf{e}}$ on site $i$ is denoted $\ket{i}$; hence, $\ket{i} = 
\ket{\mathbf{g}_1 \ldots \mathbf{e}_i \ldots \mathbf{g}_N}$. The inter-site couplings in the 
case of chromophores are given by the electrostatic dipole-dipole interaction. The coupling 
strength of a given site pair depends on the alignment of the two dipole moments, as well as 
their distance to each other. The system in the case of three sites ($N=3$) is illustrated in 
figure \ref{fig:Trimer-system}. 

A tight-binding Hamiltonian can be used to model the time evolution
of the system \cite{Renger2001}. It reads 
\begin{eqnarray}
\hat{H}^{(N)} = \sum_i^N E_{i} \ket{i}\bra{i} + 
\sum_{i\neq j}^N J_{ij}\ket{i}\bra{j} , 
\label{eq:Hamiltonian}
\end{eqnarray}
where $E_{i}$ is the excitation energy of site $\ket{i}$, i.e., the (photon) energy required to 
excite site $i$ from $\mathbf{g}$ to $\mathbf{e}$, and $J_{ij}$ is the inter-site coupling 
between $i$ and $j$. 

With the help of $\hat{H}^{(N)}$, we wish to transfer an initial excitation localized on site $1$ to 
site $N$, i.e., $\ket{1}\mapsto e^{-i\hat{H}^{(N)}t}\ket{1}=\ket{N}$ (we put $\hbar=1$ from now 
on) for some time $t=\tau$. This type of initial condition would in the case of a molecular 
aggregate correspond to a localized excitation at site $\ket{1}$ prepared by an appropriately 
tailored laser pulse. We refer to the process as {\it population transfer} and we are making 
the restriction to only one excitation at the time in the network, which reduces the 
$2^N$-dimensional Hilbert space of the system to the $N$-dimensional 
single-excitation subspace. We also restrict our study to include only real-valued coupling 
parameters. 

\section{Quantifying efficiency and coherence}
\label{sec:Quantifying-coherence-and-efficiency}
In population transfer there are two natural bases to consider; the {\it site basis}, with 
respect to which the population transfer is described, and the {\it exciton basis}, i.e., the 
Hamiltonian eigenbasis. The delocalization of the evolving quantum state over different sites 
will be governed by the Hamiltonian parameters and the relation between the coherence in 
the site and exciton bases is hence of interest. Thus, in this study we focus on these two bases 
when we are relating efficiency and coherence to each other.

\subsection{Quantifying efficiency}
Our task is to transfer an initially site-localized state $\ket{\psi (0)} = \ket{1}$ into a 
site-localized target state $\ket{\psi_{\rm{target}}} = \ket{N}$, by means of the tight-binding 
Hamiltonian in equation (\ref{eq:Hamiltonian}). To quantify the efficiency of this process, 
we use the pure state {\it fidelity} \cite{Nielsen2010}, which takes the form 
\begin{eqnarray}
F = \left| \bra{\psi_{\rm{target}}} \hat{U}(t,0) \ket{\psi (0)} \right|  = |\bra{N} \hat{U}(t,0) \ket{1}|
\end{eqnarray}
with $\hat{U}(t,0) = e^{-i\hat{H}^{(N)}t}$. Perfect population transfer corresponds to the 
case where $F=1$. 

\subsection{Quantifying coherence}
As mentioned in the introduction, the $l_{1}$-norm and the relative entropy of coherence 
(REOC) satisfy the conditions for being quantifiers of coherence \cite{Baumgratz2014}. These 
two measures have turned out to be the most frequently used when analyzing coherence in 
different processes (see, e.g., \cite{Misra2016,Lostaglio2015b}). In this study, we use both 
$l_{1}$-norm and REOC, in order to cover both geometric and entropic aspects of coherence.  

We start by introducing the free pure states as the members of an orthonormal basis 
$\psi = \{ \ket{\psi_i} \}$ spanning the Hilbert space of the system. This defines the free states 
$\hat{\sigma}$ as those that diagonalizes in $\{ \ket{\psi_i} \}$, i.e., $\hat{\sigma} = 
\sum_i \sigma_{ii} \ket{\psi_i}\bra{\psi_i}$. The idea is to use the concept of free states to 
quantify the amount of coherence relative $\psi$ in a given density operator $\hat{\rho}$. 

Our first measure is the $l_{1}$-norm of coherence, which is induced by the $l_{1}$-matrix 
norm \cite{Horn1991} as follows. Consider the difference $\bra{\psi_i} \hat{\rho} \ket{\psi_j} - 
\sigma_{ij} \equiv \rho_{ij}^{\psi} - \sigma_{ii} \delta_{ij}$. The corresponding $l_{1}$-matrix 
norm reads    
\begin{eqnarray}
D_{l_{1}}^{\psi} (\hat{\rho},\hat{\sigma}) = 
\sum_{i,j} \left|\rho_{ij}^{\psi} - \sigma_{ii} \delta_{ij} \right| . 
\end{eqnarray} 
The $l_{1}$-norm of coherence $C_{l_{1}}^{\psi}(\hat{\rho})$ of $\hat{\rho}$ with respect 
to the free basis is given by minimizing $D_{l_{1}}^{\psi} (\hat{\rho},\hat{\sigma})$ over the 
set of free states. This minimum is obtained for $\hat{\sigma} = \hat{\rho}_{{\rm {diag}}}^{\psi} = 
\sum_i \rho_{ii}^{\psi} \ket{\psi_i}\bra{\psi_i}$, yielding 
\begin{eqnarray}
C_{l_{1}}^{\psi} (\hat{\rho}) = \sum_{i\neq j} \left|\rho_{ij}^{\psi} \right| = 
2\sum_{i<j} \left| \rho_{ij}^{\psi} \right| . 
\end{eqnarray}
Note that the $l_{1}$-norm is bounded by \cite{Cheng2015}
\begin{eqnarray}
C_{l_{1}}^{\psi} (\hat{\rho})\leq d-1
\label{eq:l1bound}
\end{eqnarray}
where $d$ is the dimension of the system. The upper bound is reached for the {\it maximally 
coherent state} (MCS) \cite{Bai2015}
\begin{eqnarray}
\ket{{\rm{MCS}}} = \frac{1}{\sqrt{d}} \sum_{i=1}^d \ket{\psi_{i}}.
\label{eq:mcs}
\end{eqnarray}

Our second measure is REOC, which is defined by minimizing the relative entropy 
\begin{eqnarray}
S(\hat{\rho}\parallel\hat{\sigma})={\rm {Tr}} \left[\hat{\rho}\log_{2}\hat{\rho}\right]-{\rm {Tr}}\left[\hat{\rho}\log_{2}\hat{\sigma}\right] 
\label{eq:Relativ entropi}
\end{eqnarray}
over the set of free states $\hat{\sigma}$. This minimum can be found by using the identity 
\begin{eqnarray}
S(\hat{\rho}\parallel\hat{\sigma}) = S(\hat{\rho}_{{\rm {diag}}}^{\psi})-S(\hat{\rho}) + 
S(\hat{\rho}_{{\rm {diag}}}^{\psi} \parallel\hat{\sigma}) 
\label{eq:Relativ entropi_alt}
\end{eqnarray}
with $S(\cdot)$ the von Neumann entropy. Since the lower bound of Klein's inequality 
$S(\hat{\rho}_{{\rm {diag}}}^{\psi} \parallel\hat{\sigma}) \geq 0$ is obtained if and 
only if $\hat{\sigma} = \hat{\rho}_{{\rm {diag}}}^{\psi}$, we obtain REOC of $\hat{\rho}$
\begin{eqnarray}
C_{{\rm{REOC}}}^{\psi} (\hat{\rho})=S(\hat{\rho}_{{\rm{diag}}}^{\psi})-S(\hat{\rho}) . 
\end{eqnarray}
For pure states, $S(\hat{\rho})$ vanishes and REOC reduces to 
\begin{eqnarray}
C_{{\rm{REOC}}}^{\psi} (\hat{\rho})=S(\hat{\rho}_{{\rm {diag}}}^{\psi}).
\end{eqnarray}
REOC is bounded by 
\begin{eqnarray}
C_{\rm{REOC}}^{\psi} (\hat{\rho}) \leq \log_2 d
\label{eq:reocbound}
\end{eqnarray}
with equality for the maximally coherent state in equation (\ref{eq:mcs}). 

While $C_{l_{1}}^{\psi}$ and $C_{{\rm {REOC}}}^{\psi}$ are generally time-dependent with respect 
to the site basis ($\psi = s$) in our system, they are constant in time in the exciton basis 
($\psi = e$). The latter can be 
seen by noting that an initial state $\hat{\rho} (0) = \sum_{i,j} \rho_{ij}^{e} \ket{e_i} \bra{e_j}$, 
$\{ \ket{e_i} \}$ being the exciton basis of the time-independent Hamiltonian in equation 
(\ref{eq:Hamiltonian}), evolves into $\hat{\rho} (t) = \sum_{i,j} \rho_{ij}^{e} e^{-i\mathcal{E}_i t} 
\ket{e_i} \bra{e_j} e^{i\mathcal{E}_j t}$, where $\{ \mathcal{E}_i \}$ are the exciton energies. 
Thus, in the exciton basis 
\begin{eqnarray}
C_{l_{1}}^{e} (\hat{\rho}(t)) & = & 2\sum_{i<j} \left| e^{-i \mathcal{E}_i t} \rho_{ij}^{e} 
e^{i\mathcal{E}_{j}t} \right| = 
2\sum_{i<j} \left| \rho_{ij}^{e} \right| = C_{l_{1}}(\hat{\rho}(0))
\end{eqnarray}
for $l_1$-norm; the time independence of REOC follows immediately from that 
$\hat{\rho}_{\rm{diag}}^{e}$ is time-independent. 

\subsection{Global and local coherence}
When $\hat{\rho}$ is representing the full state of the system, $C^{s} (\hat{\rho})$ measures 
the {\it global} coherence of the network of sites. In the $N\geq 3$ case, we can also have 
{\it local} coherence for the subsystem pairs $(i,j)$. These subsystem pairs are described 
by the reduced states 
\begin{eqnarray}
\hat{\rho}_{ij} \doteq \left( \begin{array}{cccc} 1-(\rho_{ii}^{s} +\rho_{jj}^{s}) & 0 & 0 & 0\\
0 & \rho_{jj}^{s} & \rho_{ij}^{s} & 0\\
0 & \rho_{ji}^{s} & \rho_{ii}^{s} & 0\\
0 & 0 & 0 & 0
\end{array} \right) 
\end{eqnarray}
expressed in the product bases $\left\{ \ket{\mathbf{g}_i \mathbf{g}_j}, \ket{\mathbf{g}_i \mathbf{e}_j},  
\ket{\mathbf{e}_i \mathbf{g}_j}, \ket{\mathbf{e}_i \mathbf{e}_j} \right\}$. 

With respect to the product basis, the local coherence of the pair $(i,j)$ as measured 
by the $l_{1}$-norm is $C_{l_1}^{ij} (\hat{\rho}) = 2\left|  \rho_{ij}^{s} \right|$. This coincides 
with entanglement as measured by concurrence \cite{Wootters1998}, which is a consequence 
of the restriction to single-excitation subspace of the full Hilbert space of our $N$-site system. 
It is hence also possible to relate efficiency to entanglement between sites \cite{Sarovar10}.

\subsection{Comparing efficiency and coherence}
We wish to develop the quantification of coherence a bit further and specifically, we ask; how 
much coherence has there been on average in the system during population transfer? We are 
hence interested in the {\it time-averaged coherence} (denoted as TAC in the figures), defined as
\begin{eqnarray}
\overline{C}^{\psi} (t;\hat{\rho}) = \frac{1}{t} \int_0^{t} C^{\psi} (\hat{\rho}(t')) dt' , 
\end{eqnarray}
where $C^{\psi} (\hat{\rho}(t))$ is either $C_{l_{1}}^{\psi} (\hat{\rho}(t))$ 
or $C_{{\rm {REOC}}}^{\psi} (\hat{\rho}(t))$. We refer to $C^{\psi} (\hat{\rho}(t))$ as 
{\it time-local coherence} (denoted as TLC in the figures).  

In the following, we optimize $F$, $C^{\psi}$, and $\overline{C}^{\psi}$ numerically and 
analytically (whenever possible) for a dimer (section \ref{sec:dimer}) and a trimer (section 
\ref{sec:trimer}) by varying over the parameter space of the tight-binding Hamiltonian in 
equation (\ref{eq:Hamiltonian}). The coherence quantities are computed both in the site 
basis and the exciton basis as well as for both types of coherence measures ($l_{1}$-norm 
and REOC). We compare the optimal Hamiltonian parameters and time for $F$ to the optimal 
Hamiltonian parameters and time for $\overline{C}^{\psi} (t;\hat{\rho})$ to see whether they 
coincide or not, in order to analyze the role of coherence for population transfer.

\section{Dimer case}
\label{sec:dimer}  
We first consider population transfer from site $1$ to site $2$ in a dimer ($N=2$). This 
process can be understood as tunneling through an energy barrier defined by the difference 
in site energies, and driven by the inter-site coupling. In this sense, the dimer enables us 
to examine efficiency and coherence associated with the tunneling mechanism alone. 

The tight-binding dimer Hamiltonian takes the form 
\begin{eqnarray}
\hat{H}^{(2)} = \omega \big[ \cos\theta (\ket{1}\bra{1}-\ket{2}\bra{2}) + 
\sin\theta (\ket{1}\bra{2} + \ket{2}\bra{1} ) \big] , 
\label{eq:Dimer_Hamiltonian}
\end{eqnarray}
where $\omega=\sqrt{E^{2}+J_{12}^{2}}$ and $\tan\theta=J_{12}/E$, $E$ being related to 
the site excitation energies as $E =\left( E_{1}-E_{2} \right) /2$.  The time evolution 
\begin{eqnarray}
\ket{1} \mapsto \hat{U} (t,0) \ket{1} & = & 
\big( \cos \omega t - i\cos \theta \sin \omega t \big)\ket{1} 
-i\sin \theta \sin \omega t \ket{2} 
\label{eq:Dimer_tidsutveckling}
\end{eqnarray}
is characterized by the single frequency $\omega$. Note that the energy gap of the two 
exciton states is $2\omega$. Without loss of generality, we assume $E,J_{12}\geq0$ and 
$0\leq\theta\leq\frac{\pi}{2}$ (thus, $\sin\theta\geq0$ and $\cos\theta\geq0$) 
in the following.

The optimization of time-averaged coherence is performed numerically. The Hamiltonian 
parameters and time are varied over $-0.500\leq E_{i}\leq0.500$, $-0.500 \leq  J_{12} \leq 
0.500$ and $0\leq t\leq10$ in steps of $\Delta E_{i}=0.01, \Delta J_{12}=0.01$, and 
$\Delta t=0.001$, respectively. We limit the study to maximum occurring in the 
given time interval - there might be Hamiltonian parameter sets for which maximum are
obtained at $t>10$.

\subsection{Population transfer efficiency} 
The fidelity is given by
\begin{eqnarray}
F(\theta,t)=\big|\bra{2}\hat{U}(t,0)\ket{1}\big|=\sin\theta\left|\sin \omega t \right|,
\label{eq:Dimer_fidelitet}
\end{eqnarray}
which is a periodic function in time with period (revival time) $T=\pi/\omega$. The fidelity 
reaches its maximum 
\begin{eqnarray} 
F_{\max} (\theta) = \max_t F(\theta,t) = \sin \theta 
\label{eq:maxfidelity}
\end{eqnarray} 
at $t = \pi/(2\omega) = \tau$, being half the revival time. Thus, the speed of the transfer process 
is inversely proportional to the energy gap $2\omega$. Perfect population transfer, i.e., $F=1$,  
occurs for $\theta = \pi /2$, which holds whenever $E=0$ and $J_{12}\neq0$. 

\subsection{Coherence}
The $l_1$-norm and REOC are directly related for pure dimer states $\hat{\rho} = \ket{\phi} 
\bra{\phi}$. To see this, we express $\ket{\phi}$ in terms of an arbitrary free basis 
$\psi = \{ \ket{\psi_1},\ket{\psi_2} \}$ as $\ket{\phi} = c_1 \ket{\psi_1} + c_2 \ket{\psi_2}$, 
yielding 
\begin{eqnarray} 
C_{l_1}^{\psi} (\hat{\rho}) = 2\left| c_1 c_2 \right| = 2\left| c_2 \right| \sqrt{1-\left| c_2 \right|^2}
\label{eq:l1_dimer_general}
\end{eqnarray}  
and 
\begin{eqnarray} 
C_{\rm{REOC}}^{\psi} (\hat{\rho} ) = - \left| c_1 \right|^2 \log_2 \left| c_1 \right|^2 - 
\left| c_2 \right|^2 \log_2 \left| c_2 \right|^2 = h\left( \left| c_2 \right|^2 \right) , 
\label{eq:reoc_dimer_general}
\end{eqnarray}  
where $h(x)=-(1-x)\log_{2}(1-x)-x\log_{2}x$ is the binary Shannon entropy and we have 
used that $\left| c_1 \right|^2 + \left| c_2 \right|^2 = 1$. By inverting equation 
(\ref{eq:l1_dimer_general}) and inserting into equation (\ref{eq:reoc_dimer_general}), we find 
\begin{eqnarray}
C_{\rm{REOC}}^{\psi} (\hat{\rho}) = 
h\left( \frac{1+\sqrt{1-\left[ C_{l_1}^{\psi} (\hat{\rho}) \right]^2}}{2} \right) , 
\end{eqnarray}
which establishes a general relation between the two coherence measures in the dimer case. We 
further note that $C^{\psi} \leq 1$ with equality for $\ket{\phi} = 
\ket{\rm{MCS}} = \frac{1}{\sqrt{2}} \left( \ket{\psi_1} + \ket{\psi_2} \right)$. 

\subsection{Comparing population transfer efficiency and coherence}

\paragraph{Site basis}
The time-local $l_{1}$-norm and REOC in the site basis are given by
\begin{eqnarray}
C_{l_{1}}^{s} (\theta,t) & = & 2\sqrt{1-\sin^{2}\theta\sin^2 \omega t}\sin\theta|\sin \omega t| 
\nonumber \\ 
 & = & 2F(\theta,t)\sqrt{1-\left[F(\theta,t)\right]^{2}} 
\label{eq:Dimer_site_l1-norm_fidelity}
\end{eqnarray}
and
\begin{eqnarray}
C_{{\rm{REOC}}}^{s} (\theta,t) & = & h(\sin^{2}\theta\sin^{2} \omega t) = 
h\left( \left[ F(\theta,t) \right]^2 \right) ,  
\label{eq:Dimer_site_REOC_fidelity}
\end{eqnarray}  
where we have used the expression for the fidelity in equation (\ref{eq:Dimer_fidelitet}). 
We see that $C_{l_{1}}^{s} (\theta,t)$ and $C_{{\rm {REOC}}}^{s} (\theta,t)$ take 
their maximal value at $F(\theta,t) = \sin \theta$ for $\sin \theta \leq \frac{1}{\sqrt{2}}$ 
and at $F(\theta,t)=\frac{1}{\sqrt{2}}$ for $\sin \theta \geq \frac{1}{\sqrt{2}}$, where the former 
occurs at the time of maximal population transfer $t = \pi /(2\omega)$ and the latter at 
\begin{eqnarray} 
t = \frac{1}{\omega} \arcsin \left[ \frac{1}{\sqrt{2}\sin \theta} \right] . 
\label{eq:maxcohtime}
\end{eqnarray} 
Explicitly, these maximal coherence values are 
\begin{eqnarray}
\max_t C_{l_1}^{s} (\theta,t) 
 & = & \left\{ \begin{array}{cc} 
\sin 2\theta , & 0 \leq \theta \leq \frac{\pi}{4} ,  \\ 
1, & \frac{\pi}{4} \leq \theta \leq \frac{\pi}{2} , 
\end{array} \right. 
\nonumber \\ 
\max_t C_{\rm{REOC}}^{s} (\theta,t) 
 & = & \left\{ \begin{array}{cc} 
h\left( \sin^2 \theta \right) , & 0 \leq \theta \leq \frac{\pi}{4} ,  \\ 
1, & \frac{\pi}{4} \leq \theta \leq \frac{\pi}{2} .
\end{array} \right.
\end{eqnarray} 

We note that while maximal efficiency and maximal coherence occur simultaneously for 
$0 \leq \theta \leq \frac{\pi}{4}$, maximal coherence precedes maximal population transfer 
for $\frac{\pi}{4} < \theta \leq \frac{\pi}{2}$, as is evident from equation (\ref{eq:maxcohtime}). 
Furthermore, when the efficiency increases, an increasingly larger part of the coherence is 
localized in time before the transfer has been completed. This indicates that time-local 
coherence plays a role for the population transfer in the dimer system. 
 
\begin{figure}[htb!]
\centering
\includegraphics[width=0.47\textwidth]{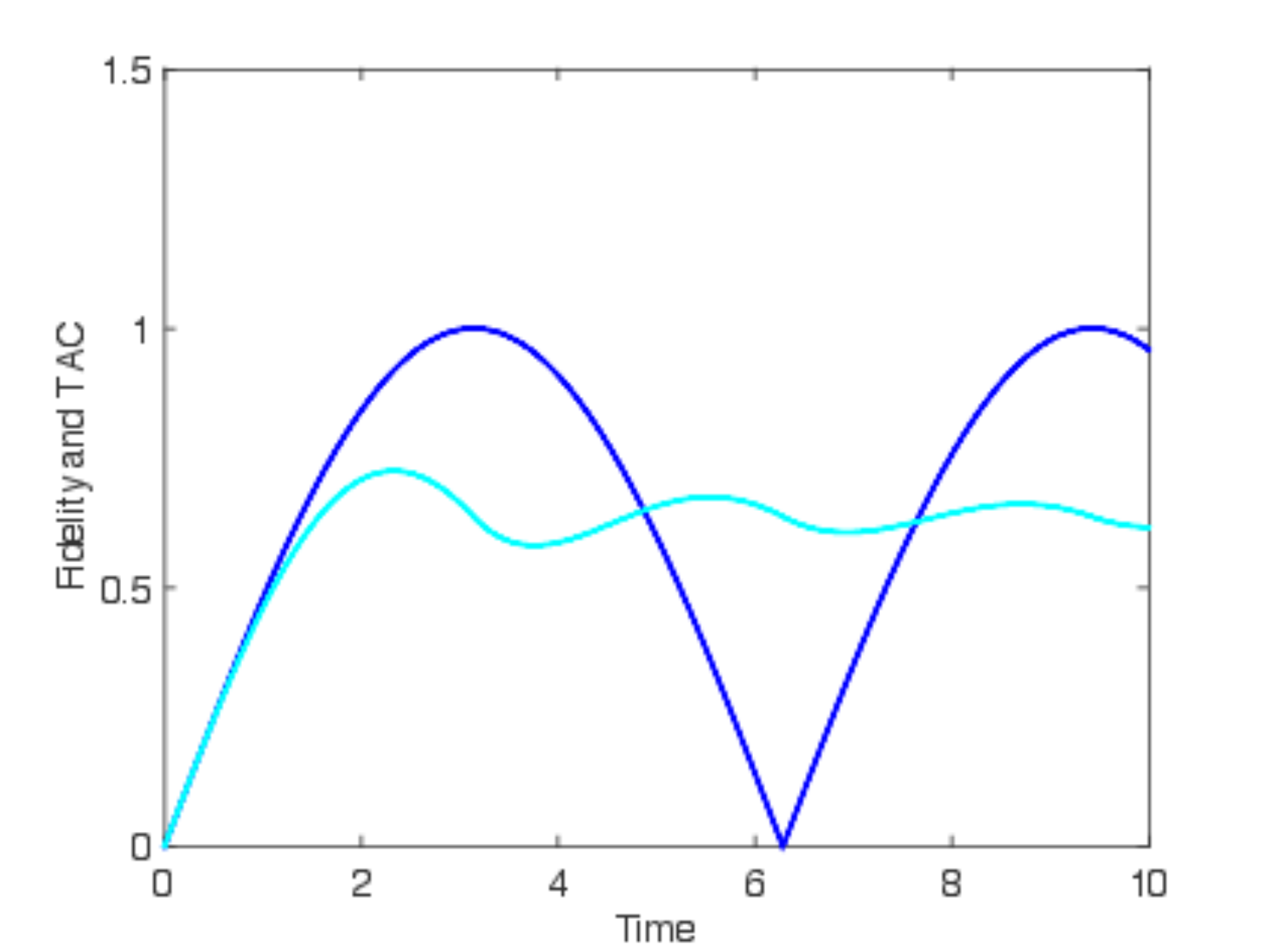}\phantom{.}
\includegraphics[width=0.47\textwidth]{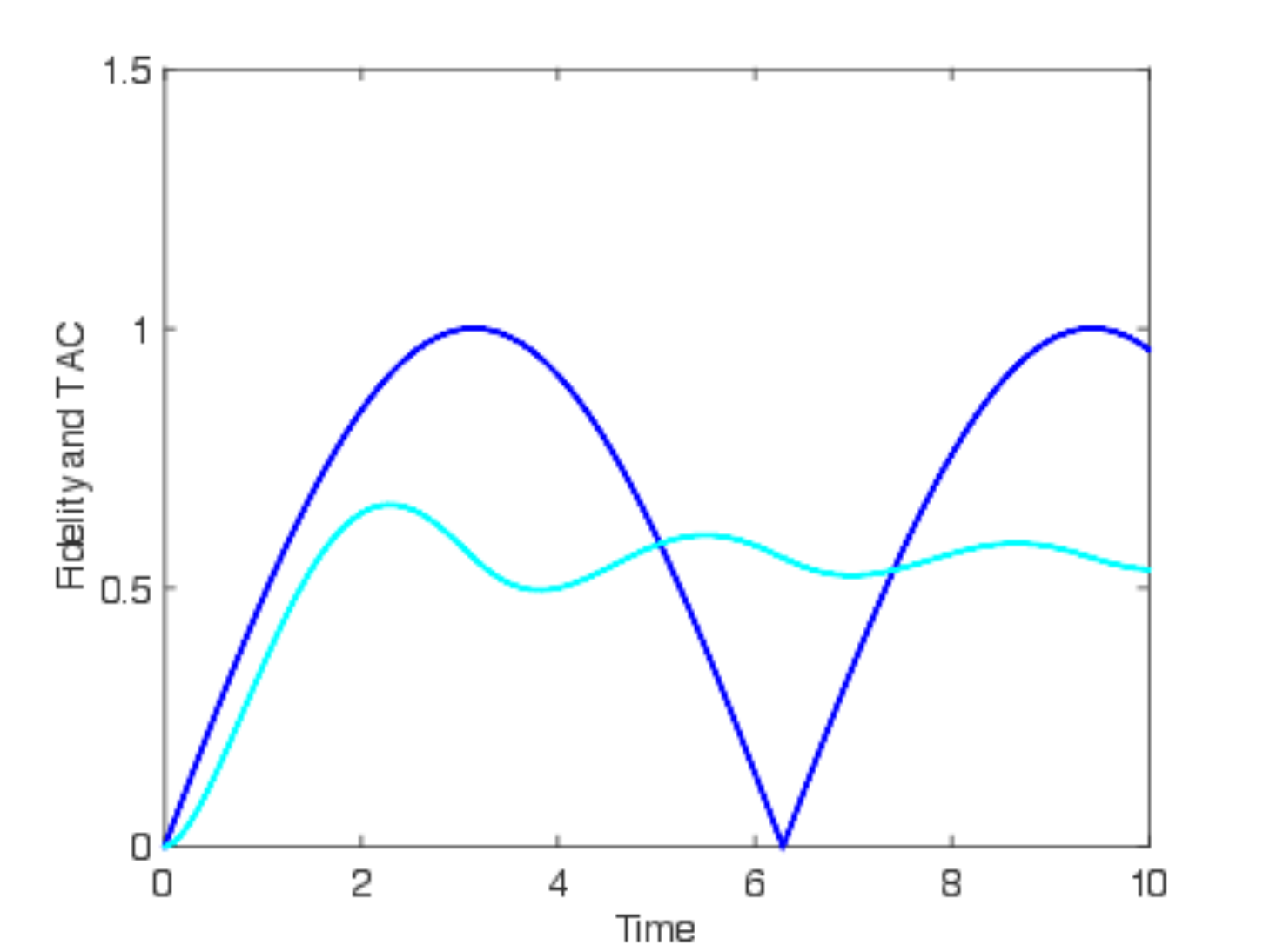}\phantom{.}
\caption{Efficiency and time-averaged coherence for a Hamiltonian that optimizes $F(\theta,\tau)$ 
in a dimer. The left panel shows $F(\theta,t)$ (blue) and $\overline{C}_{l_1}^s(\theta,t)$ (cyan).  
The right panel shows $F(\theta,t)$ (blue) and $\overline{C}_{\rm{REOC}}^s(\theta,t)$ (cyan).}
\label{fig:Dimer_site_F_vs_AC_F-max}
\end{figure}

\begin{figure}[htb!]
\centering
\includegraphics[width=0.47\textwidth]{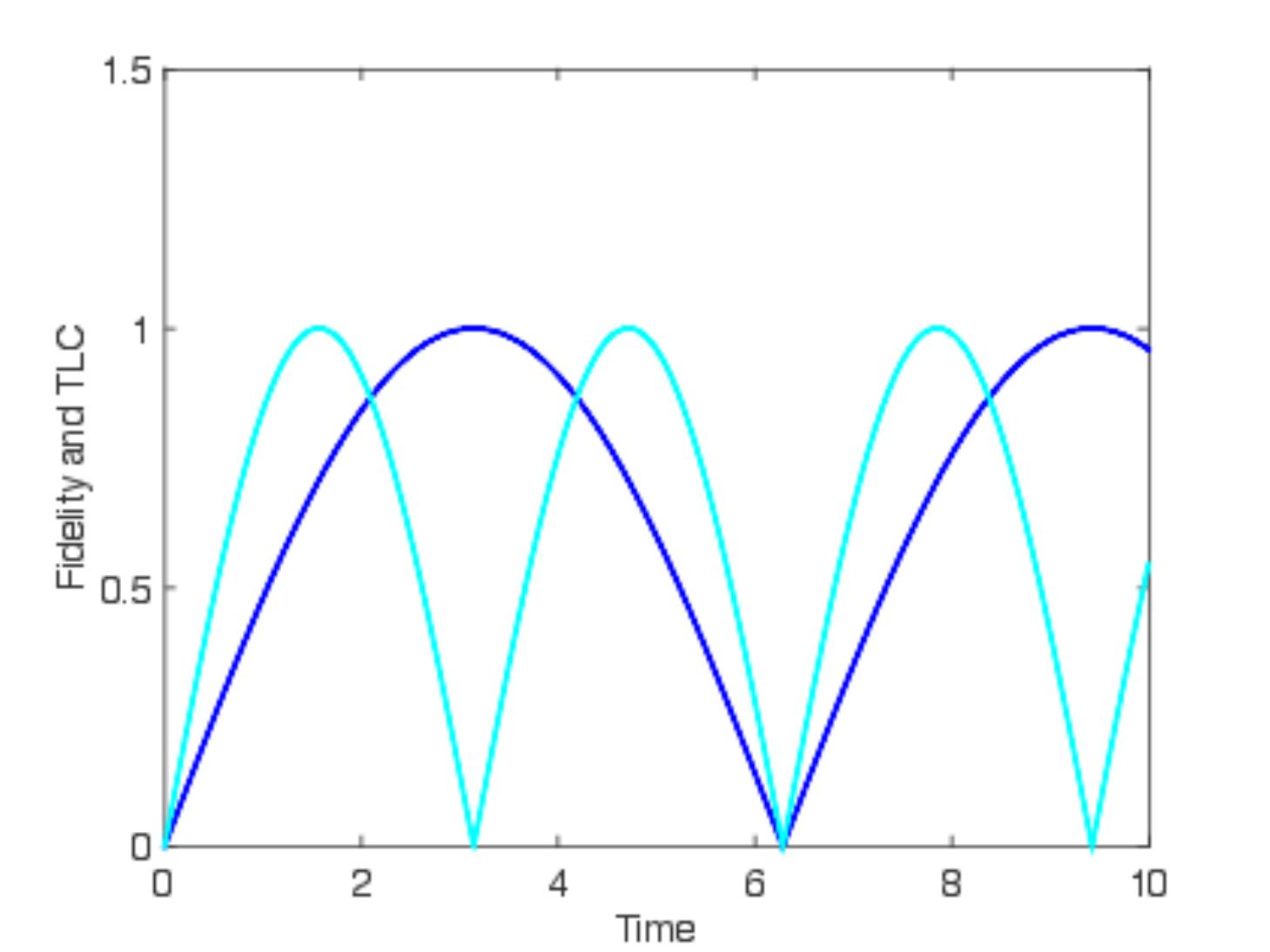}\phantom{.}
\includegraphics[width=0.47\textwidth]{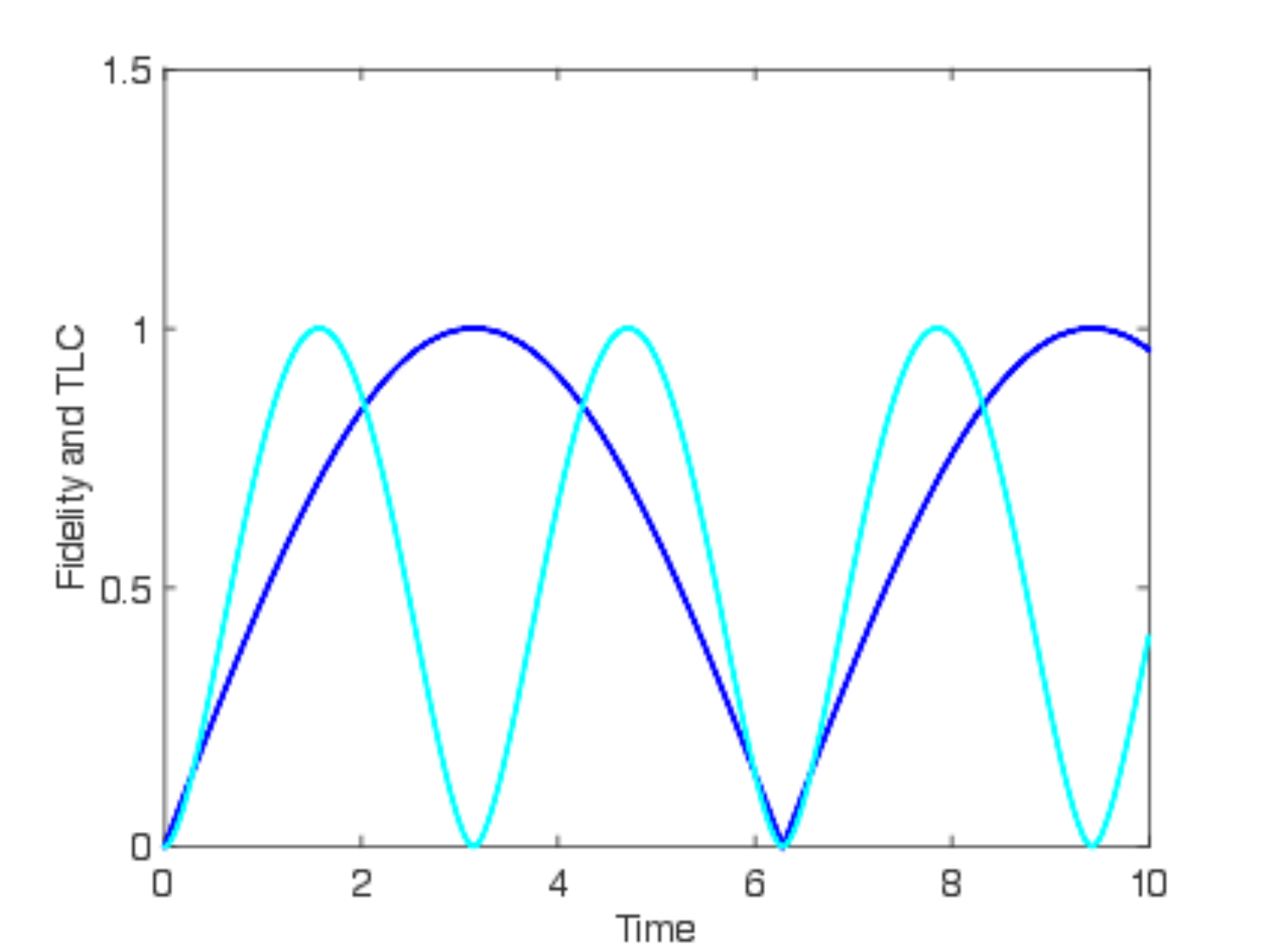}\phantom{.}
\caption{Efficiency and time-local coherence for a Hamiltonian that optimizes 
$F(\theta,\tau)$ in a dimer. The left panel shows $F(\theta,t)$ (blue) and $C_{l_1}^s(\theta,t)$ 
(cyan). The right panel shows $F(\theta,t)$ (blue) and $C_{\rm{REOC}}^s(\theta,t)$ (cyan).}
\label{fig:Dimer_site_F_vs_C_F_max}
\end{figure}

The relationship between $F(\theta,t)$ and $\overline{C}^s(\theta,t)$ and between $F(\theta,t)$ 
and $C^s(\theta,t)$ for perfect population transfer $\theta = \pi/2$, i.e., $E=0$, can be seen 
in figures \ref{fig:Dimer_site_F_vs_AC_F-max} and \ref{fig:Dimer_site_F_vs_C_F_max}, respectively. 
The results are shown for $J_{12}=\frac{1}{2}$ for which $\tau=\pi$. A closer inspection 
reveals that $\overline{C}^s(\theta,t)$ in figure \ref{fig:Dimer_site_F_vs_AC_F-max} oscillates 
roughly around the time-averaged coherence $\overline{C}^s(\theta,\tau)$. The amplitude 
of these oscillations decreases, suggesting the existence of a long-time asymptotic value, 
as will be discussed in detail in section \ref{sec:long-term} below. The time-local coherence  
in figure \ref{fig:Dimer_site_F_vs_C_F_max} maximize at precisely half the transfer time, 
in accordance with equation (\ref{eq:maxcohtime}); on the other hand, they vanish when 
the perfect transfer is completed. In terms of correlation between the two sites, 
efficiency is improved when entanglement is built up in a symmetric fashion to its 
maximum at exactly half the population transfer time and then drops to zero when the 
population returns to site $1$.

\begin{figure}[htb!]
\centering
\includegraphics[width=0.47\textwidth]{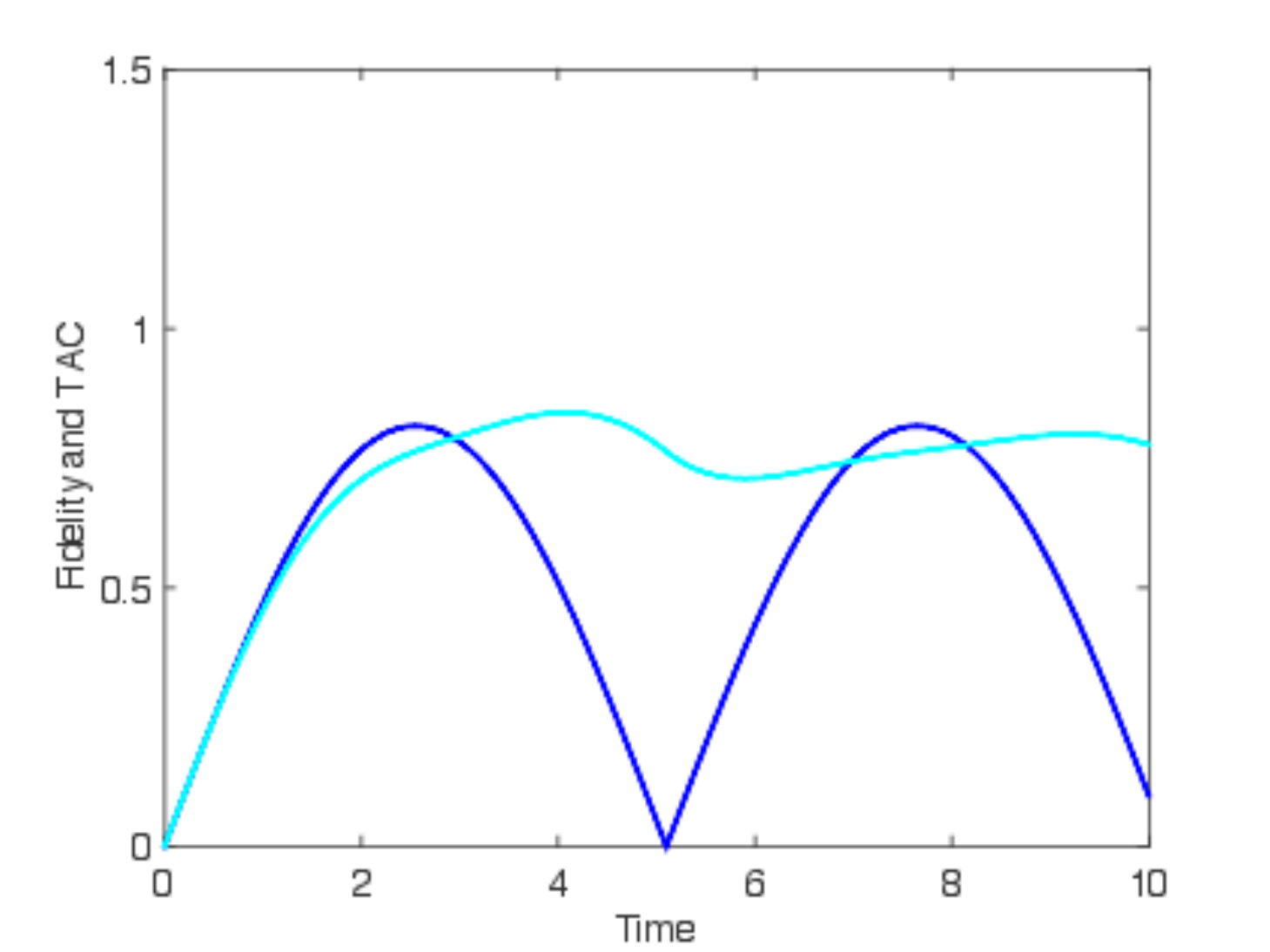}\phantom{.}
\includegraphics[width=0.47\textwidth]{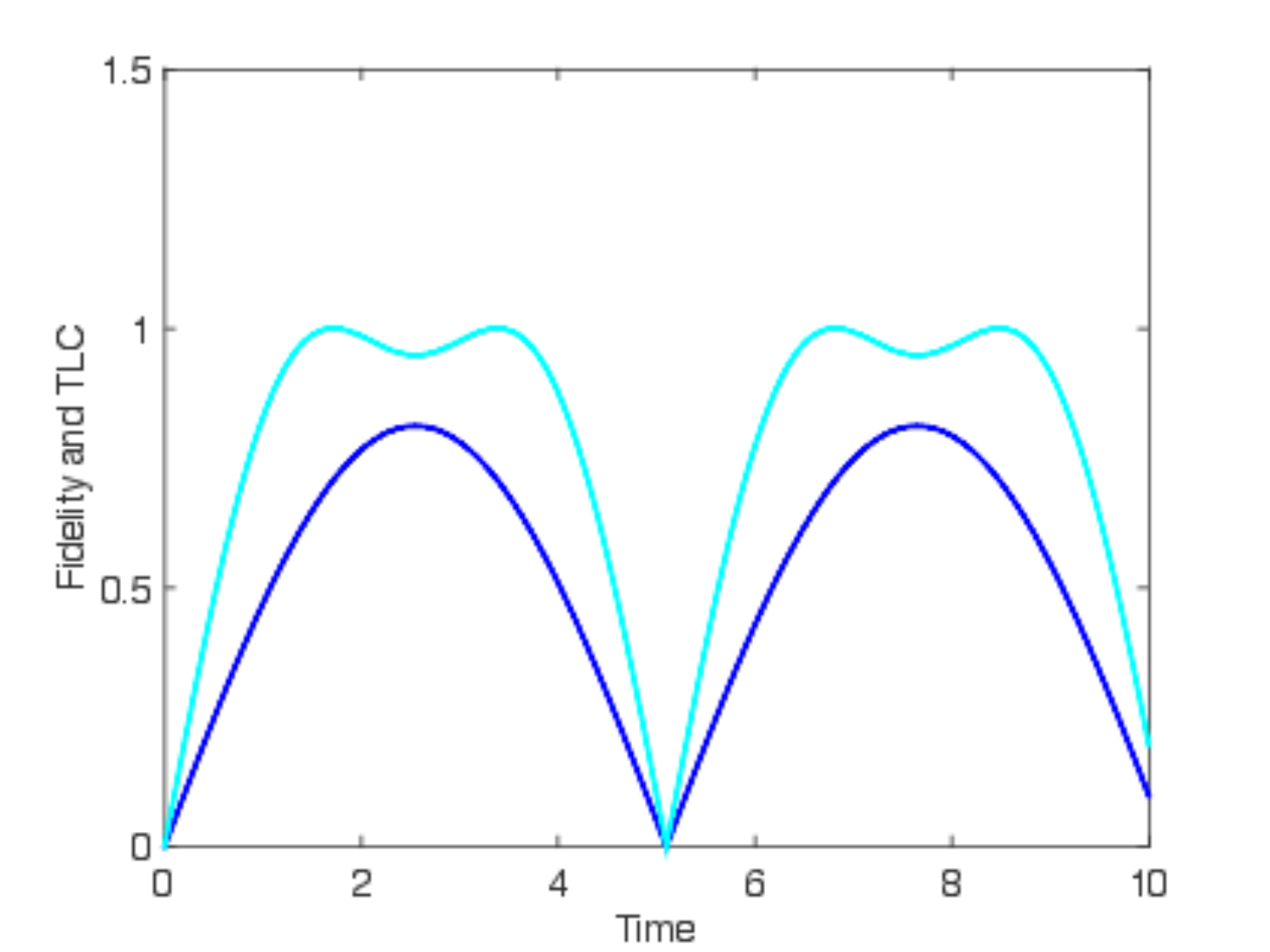}\phantom{.}
\caption{Efficiency and coherence for a Hamiltonian that optimizes $\overline{C}_{l_{1}}^s 
(\theta,t)$ in a dimer. The left panel shows $F(\theta,t)$ (blue) and 
$\overline{C}_{l_{1}}^s(\theta,t)$ (cyan). The right panel shows $F(\theta,t)$ (blue) and 
$C_{l_{1}}^s(\theta,t)$ (cyan).}
\label{fig:Dimer_site_F_vs_AC-l1_AC-l1_max}
\end{figure}

\begin{figure}[htb!]
\centering
\includegraphics[width=0.47\textwidth]{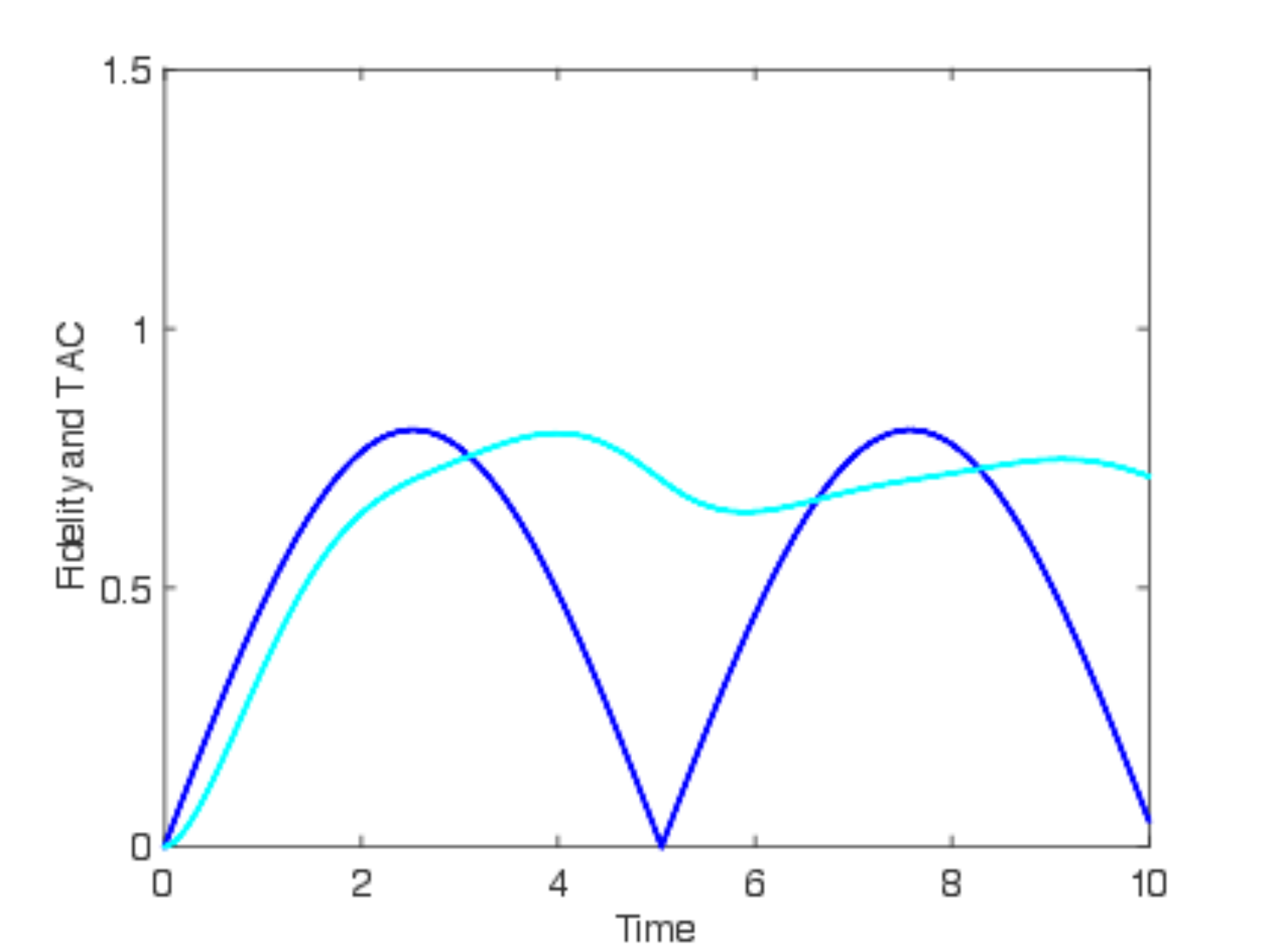}\phantom{.}
\includegraphics[width=0.47\textwidth]{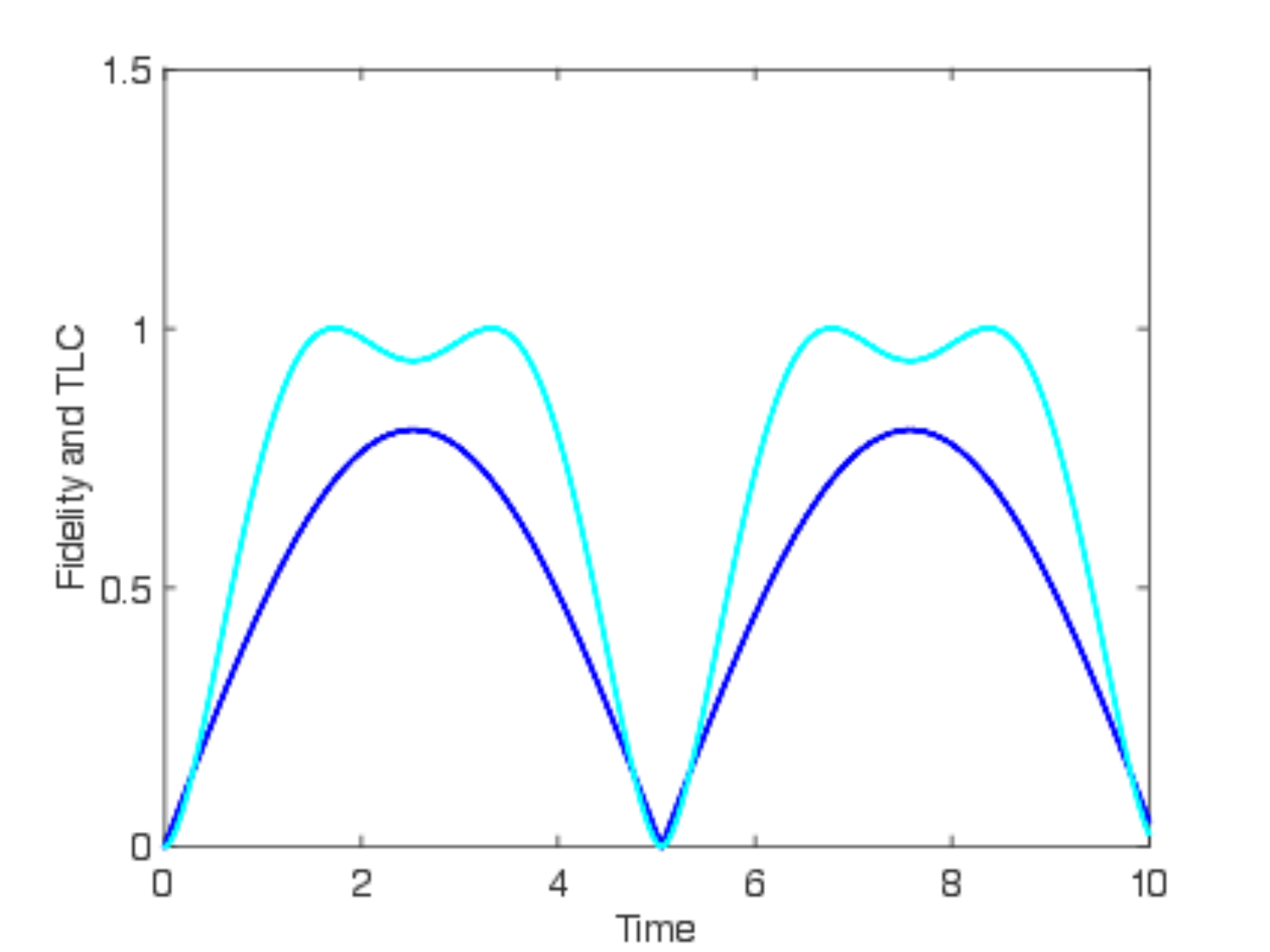}\phantom{.}
\caption{Efficiency and coherence for a Hamiltonian that optimizes $\overline{C}_{\rm{REOC}}^s 
(\theta,t)$ in a dimer. The left panel shows $F(\theta,t)$ (blue) and 
$\overline{C}_{\rm{REOC}}^s(\theta,t)$ (cyan). The right panel shows $F(\theta,t)$ (blue) 
and $C_{\rm{REOC}}^s(\theta,t)$ (cyan).}
\label{fig:Dimer_site_F_vs_AC_re_AC_re_max}
\end{figure}

The relationship between $F(\theta,t)$ and $\overline{C}^s(\theta,t)$ and between 
$F(\theta,t)$ and $C^s(\theta,t)$ 
for parameters optimizing the time-averaged coherence $\overline{C}^s(\theta,t)$, 
can be seen in figures \ref{fig:Dimer_site_F_vs_AC-l1_AC-l1_max} ($l_{1}$-norm) and 
\ref{fig:Dimer_site_F_vs_AC_re_AC_re_max} (REOC). The Hamiltonian parameters, shown 
in table \ref{tab:Dimer_tabell}, differ slightly whether the optimization is for time-averaged 
$l_1$-norm or REOC. As can be seen, perfect population transfer can never occur for these 
parameter sets, i.e., maximal time-averaged coherence does not imply maximal efficiency 
in the population transfer. We further note that $F(\theta,t)$ and $\overline{C}^s(\theta,t)$ 
do not maximize simultaneously and that $C^s(\theta,t)$ is nonzero at maximal transfer 
for these two parameter sets. Thus, the efficiency and time-averaged coherence order the 
Hamiltonians differently with respect to their ability to transfer population between sites. 

\begin{table}
\centering
\begin{tabular}{|c|c|c|}
\hline 
 & $E$ & $J_{12}$  \tabularnewline
\hline 
\hline 
$l_{1}$-norm & 0.36 & 0.50 \tabularnewline
\hline 
REOC & 0.37 & 0.50 \tabularnewline
\hline 
\end{tabular}\protect\caption{\label{tab:Dimer_tabell} Hamiltonian parameters optimizing 
$\overline{C}^s$ in a dimer. These parameter values correspond to $F_{\max}(\theta) 
\approx 0.81$ and $0.80$ for optimal $l_1$-norm and REOC, respectively.}
\end{table}

\paragraph{Exciton basis}

The exciton states read 
\begin{eqnarray}
\ket{e_{+}} & = & \cos\frac{\theta}{2}\ket{1}+\sin\frac{\theta}{2}\ket{2}, 
\nonumber \\
\ket{e_{-}} & = & -\sin\frac{\theta}{2}\ket{1}+\cos\frac{\theta}{2}\ket{2}
\end{eqnarray}
with exciton energies $\mathcal{E}_{\pm} = \pm\omega$. We find the $l_{1}$-norm 
measure of coherence and REOC to be
\begin{eqnarray}
C_{l_{1}}^{e} (\theta) = \sin\theta = F_{\max} (\theta) 
\label{eq:Dimer_exciton_l1-norm}
\end{eqnarray}
and 
\begin{eqnarray}
C_{{\rm {REOC}}}^{e} (\theta) = h\left(\frac{1+\sqrt{1-\sin^{2}\theta}}{2}\right) = 
h\left(\frac{1+\sqrt{1-\left[F_{\max} (\theta) \right]^2}}{2}\right) ,  
\label{eq:Dimer_exciton_REOC}
\end{eqnarray}
respectively, where we have used the expression for the fidelity maximum in equation 
(\ref{eq:maxfidelity}). Since coherence is time-independent in the exciton basis, there is apparently 
no distinction between the time-local and time-averaged coherence in this basis.

The key observation is that $C_{l_{1}}^{e} (\theta)$ and $C_{{\rm {REOC}}}^{e} (\theta)$
are strictly monotonically increasing functions of the fidelity maximum. This implies that coherence 
with respect to the exciton basis increases with the population transfer efficiency and becomes 
maximal for perfect population transfer. Thus, population transfer efficiency is in one-to-one 
correspondence with coherence in the exciton basis in the case of a dimer system.

\section{Trimer case}
\label{sec:trimer}
We next present our results for the case of population transfer from site $1$ to site $3$ in 
a trimer. This process can take place via two distinct pathways, either $1 \rightarrow 3$ 
directly or $1 \rightarrow 2 \rightarrow 3$. In this sense, the trimer is the smallest system 
that can capture constructive and destructive interference between different pathways in a site 
network. 

The tight-binding trimer Hamiltonian reads 
\begin{eqnarray}
\hat{H}^{(3)} & = & E_{1}\ket{1}\bra{1}+E_{2}\ket{2}\bra{2}+E_{3}\ket{3}\bra{3} 
\nonumber \\
 & & + (J_{12}\ket{1}\bra{2}+J_{23}\ket{2}\bra{3}+J_{13}\ket{1}\bra{3}+{\rm {h.c.}}).
\label{eq:Trimer_Hamiltonian}
\end{eqnarray}

Optimization of population transfer efficiency and time-averaged coherence is performed 
numerically. The Hamiltonian parameters and time are varied over 
$-0.500 \leq E_i \leq 0.500$, $-0.500 \leq  J_{ij} \leq 0.500$ and $0\leq t\leq10$ 
in steps of $\Delta E_i = 0.1, \Delta J_{ij}=0.1$, and $\Delta t=0.01$, respectively. The 
larger step size in the trimer compared to the dimer is due to computational limitations. 

It turns out again that optimal efficiency in the 
population transfer is not obtained for parameter values that optimize the time-averaged 
coherence. In the trimer, unlike in the dimer, the sign of the relative site energies and 
inter-site couplings matters; two parameter sets only differing in the signs of these parameters
have (in general) different maxima for fidelity and time-averaged coherence. By using the bounds in 
equations (\ref{eq:l1bound}) and (\ref{eq:reocbound}), we find 
\begin{eqnarray}
C_{l_{1}}^{\psi} (\hat{\rho}) \leq 2
\end{eqnarray}
and
\begin{eqnarray}
C_{\rm{REOC}}^{\psi} (\hat{\rho}) \leq \log_2 3 \approx 1.59
\end{eqnarray}
with equality for the maximally coherent state $\ket{\rm{MCS}} = \frac{1}{\sqrt{3}} \sum_{i=1}^{3}\ket{\psi_{i}}$. 

\subsection{Efficiency and coherence for perfect population transfer}
To put the Hamiltonian on convenient form, we note that the local energy term 
$\sum_{i=1}^{3} E_{i}\ket{i}\bra{i}$ in equation (\ref{eq:Trimer_Hamiltonian}) can be 
rewritten as 
\begin{eqnarray}
\sum_{i=1}^{3}E_{i}\ket{i}\bra{i} & \sim & 
\frac{E_{1}-E_{3}}{2} \Big( \ket{1}\bra{1}-\ket{3}\bra{3} \Big) + 
\left(E_{2} - \frac{E_{1}+E_{3}}{2} \right) \ket{2}\bra{2} 
\nonumber \\ 
 & = & E_{13} \Big( \ket{1}\bra{1}-\ket{3}\bra{3} \Big) + 
\widetilde{E} \ket{2}\bra{2} , 
\label{eq:localenergytrimer}
\end{eqnarray}
where we have ignored a trivial zero-point energy term and defined $E_{13} = (E_1-E_3)/2$ 
and $\widetilde{E} = E_2 - (E_1 + E_3)/2$. The results from the numerical calculations show 
that a necessary, but not sufficient condition for perfect population transfer between sites $1$ 
and $3$ is $E_{13}=0$ and $\left|J_{12}\right| = \left|J_{23}\right|$, the latter being 
consistent with Lemma 2 of \cite{Kay2010} for an open spin chain. By combining these 
observations and equation (\ref{eq:localenergytrimer}), we may put $J_{12}=\sigma J_{23}$ 
with the `parity' $\sigma=\pm1$ in terms of which the trimer Hamiltonian takes the form  
\begin{eqnarray}
\hat{H}^{(3)} & = & \widetilde{E}\ket{2}\bra{2}+J_{23}\Big(\ket{1}\bra{2} + 
\ket{2}\bra{1}\Big) 
\nonumber \\
 & & + \sigma J_{23} \Big( \ket{2} \bra{3} + \ket{3}\bra{2} \Big) + J_{13} \Big( \ket{1}\bra{3} + 
\ket{3}\bra{1} \Big)
\label{eq:Trimer_Hamiltonian_F_max}
\end{eqnarray}
when it allows for perfect population transfer. Here, $J_{12}=\sigma J_{23}$ with the 
`parity' $\sigma=\pm1$. By inspection, we find an exciton state
\begin{eqnarray}
\ket{e_d} = \frac{1}{\sqrt{2}} \big( \ket{1}-\sigma\ket{3} \big)
\end{eqnarray}
with corresponding exciton energy $\mathcal{E}_d = -\sigma J_{13}$. Note that 
$\ket{e_d}$ is a `dark' eigenstate in the sense that it does not contain the state $\ket{2}$. 
In order to find the remaining two exciton states and energies, we eliminate $\ket{e_d}$ and 
find the $2\times2$ block 
\begin{eqnarray}
H_{2\times2}^{(3)} = \widetilde{E}\ket{2}\bra{2}+\sigma J_{13}\ket{b}\bra{b} 
+\sqrt{2}J_{23}\Big(\ket{2}\bra{b}+\ket{b}\bra{2}\Big) 
\end{eqnarray}
with $\ket{b}=\frac{1}{\sqrt{2}}\big(\ket{1}+\sigma\ket{3}\big)$. This can be diagonalized 
by standard methods, yielding the exciton energies
\begin{eqnarray}
\mathcal{E}_{\pm} = \frac{\widetilde{E}+\sigma J_{13}}{2} \pm \frac{1}{2} 
\sqrt{(\widetilde{E}-\sigma J_{13})^{2}+8J_{23}^{2}}
\end{eqnarray}
and corresponding exciton states 
\begin{eqnarray}
\ket{e_{+}} & = & \cos\frac{\vartheta}{2}\ket{2}+\sin\frac{\vartheta}{2}\ket{b}, 
\nonumber \\
\ket{e_{-}} & = & -\sin\frac{\vartheta}{2}\ket{2}+\cos\frac{\vartheta_{}}{2}\ket{b},
\end{eqnarray}
where
\begin{eqnarray}
\tan\vartheta = \frac{2\sqrt{2}J_{23}}{\widetilde{E}-\sigma J_{13}}.
\end{eqnarray}
We note that the exciton states and energies depend nontrivially on the parity $\sigma$. 
Thus, the two parity cases must be treated separately in the following analysis.

The time evolution operator associated with the Hamiltonian in equation 
(\ref{eq:Trimer_Hamiltonian_F_max}) takes the form
\begin{eqnarray}
\hat{U}(t,0) & = & e^{-i\mathcal{E}_d t}\ket{e_d}\bra{e_d} + 
e^{-i\mathcal{E}_{+}t} \ket{e_{+}}\bra{e_{+}} 
\nonumber \\
 &  & +e^{-i\mathcal{E}_{-}t}\ket{e_{-}}\bra{e_{-}}.
\end{eqnarray}
We find the fidelity of the population transfer $1\rightarrow3$: 
\begin{eqnarray}
F(\vec{h},t) & = & |\bra{3}\hat{U}(t,0)\ket{1}| 
\nonumber \\
 & = & \frac{1}{2}\left|1-e^{-i\left(\mathcal{E}_{+}-\mathcal{E}_d \right)t}\sin^{2}\frac{\vartheta}{2} - 
e^{-i\left(\mathcal{E}_{-}-\mathcal{E}_d \right)t}\cos^{2}\frac{\vartheta}{2}\right| 
\nonumber \\
 & = & \frac{1}{2}\left|1-e^{-i\omega_{+} t}\sin^2 \frac{\vartheta}{2} - 
e^{-i\omega_{-}t} \cos^2 \frac{\vartheta}{2}\right|,
\label{eq:Trimer_fidelitet}
\end{eqnarray}
where $\vec{h}$ denote the set of relevant Hamiltonian parameters and we have 
identified two fundamental frequencies
\begin{eqnarray} 
\omega_{\pm} & = & \mathcal{E}_{\pm}-\mathcal{E}_d = 
\frac{\widetilde{E}+3\sigma J_{13}}{2}\pm\frac{1}{2} 
\sqrt{(\widetilde{E}-\sigma J_{13})^{2}+8J_{23}^{2}} .
\label{eq:Trimer_frekvenser}
\end{eqnarray} 
Perfect population transfer occurs provided there exists a `perfect transfer 
time' $\tau$ satisfying
\begin{eqnarray}
\omega_{\pm} \tau = (2n_{\pm} + 1) \pi,
\end{eqnarray}
where $n_{\pm}$ are arbitrary integers. We thus end up with the following necessary and 
sufficient condition: perfect population transfer occurs if and only if there exist integers 
$n_{\pm}$ such that
\begin{eqnarray}
\frac{\omega_{+}}{\omega_{-}}=\frac{2n_{+}+1}{2n_{-}+1}.
\end{eqnarray}

Let us consider the two special cases $\omega_{+} = \pm \omega_{-}$. By using equation 
(\ref{eq:Trimer_frekvenser}), we see that the $\omega_{+} = \omega_{-}$ case corresponds 
$\widetilde{E}=\sigma J_{13}$ and $J_{23}=0$. Here, the Hamiltonian takes the form
\begin{eqnarray}
\hat{H}^{(3)}=J_{13}\Big(\sigma\ket{2}\bra{2}+\ket{1}\bra{3}+\ket{3}\bra{1}\Big)
\end{eqnarray}
and the smallest positive perfect transfer time is
\begin{eqnarray}
\tau=\frac{\pi}{2|J_{13}|}.
\end{eqnarray}
Similarly, the $\omega_{+} = - \omega_{-}$ case corresponds $\widetilde{E}+3\sigma J_{13}=0$
and $J_{23}$ arbitrary. The corresponding Hamiltonian reads
\begin{eqnarray}
\hat{H}^{(3)} & = & -3\sigma J_{13}\ket{2}\bra{2}+J_{23}\Big(\ket{1}\bra{2}+\ket{2}\bra{1}\Big)
\nonumber \\
 & & +\sigma J_{23}\Big(\ket{2}\bra{3}+\ket{3}\bra{2}\Big)+J_{13} 
\Big(\ket{1}\bra{3}+\ket{3}\bra{1}\Big),
\label{eq:Trimer_Lambda_system}
\end{eqnarray}
which contains for $J_{13}=0$ a resonant $\Lambda$ system \cite{Hu2000} with real-valued 
and equal inter-site couplings. The smallest positive transfer time now reads
\begin{eqnarray}
\tau=\frac{\pi}{\sqrt{4J_{13}^{2}+2J_{23}^{2}}}.
\end{eqnarray}

\paragraph{Coherence in the site basis}
The $l_{1}$-norm in the site basis is determined by the off-diagonal elements 
\begin{eqnarray}
\left|\rho_{ij}^{s} (t) \right| & = & 
\left|\bra{i}\hat{U}(t,0) \ket{1}\bra{1} U^{\dagger} (t,0) \ket{j}\right| 
\nonumber \\
 & \equiv & f_i (t)f_j (t),
\end{eqnarray}
where $f_i (t)=\left|\bra{i}\hat{U}(t,0)\ket{1}\right|$ and we have used that 
$\bra{1}U^{\dagger}(t,0)\ket{j}=\bra{j}\hat{U}(t,0)\ket{1}^{\ast}$. Physically, $[f_1 (t)]^{2}$ 
is the survival probability of the initial state; $f_2 (t)$ ($f_3 (t)$) is the fidelity 
of the $1\rightarrow2$ ($1\rightarrow3$) population transfer. We find the $l_{1}$-norm
\begin{eqnarray}
C_{l_{1}}^{s} (\vec{h},t) = 2\big[ f_{1}(t)f_{2}(t)+f_{1}(t)f_{3}(t)+f_{2}(t)f_{3}(t) \big] .
\end{eqnarray}
Similarly, REOC is determined by the incoherent density operator
\begin{eqnarray}
\hat{\rho}_{\rm{diag}}^s (t)=\sum_{i=1}^{3}\left[ f_{i} (t) \right]^{2} \ket{i} \bra{i}.
\end{eqnarray}
Explictly,
\begin{eqnarray}
C_{\rm{REOC}}^{s} (\vec{h},t) = -\sum_{i=1}^{2}\left[ f_{i}(t) \right]^{2} \log_{2} \left[ f_i (t) \right]^{2}.
\end{eqnarray}
We see that the $l_1$-norm and REOC in the trimer are determined by the same three 
functions $f_1(t),f_2(t)$, and $f_3(t)$. 

For $E_{13}=0$ and $|J_{12}|=|J_{23}|$ we have that $f_{3}(t)=F(\vec{h},t)$ 
given by equation (\ref{eq:Trimer_fidelitet}). The remaining two functions read 
\begin{eqnarray}
f_1 (t) & = & \frac{1}{2}\left|1+e^{-i\omega_{+} t}\sin^{2}\frac{\vartheta}{2} + 
e^{-i\omega_{-}t}\cos^{2}\frac{\vartheta}{2}\right|,
\nonumber \\
f_2 (t) & = & \frac{1}{\sqrt{2}} \left| \sin\vartheta 
\sin \left[ \frac{\left(\omega_{+}-\omega_{-} \right)t}{2} \right] \right|.
\end{eqnarray} 

\paragraph{Exciton basis}
Since the coherence in the exciton basis are time-independent, all information is contained 
in the initial state $\ket{1}$ expressed in terms of the exciton states. Explicitly, we find 
\begin{eqnarray}
\ket{1} = \frac{1}{\sqrt{2}} \left( \ket{e_d}+\sin\frac{\vartheta}{2}\ket{e_{+}} + 
\cos\frac{\vartheta}{2}\ket{e_{-}} \right).
\label{eq:Trimer_exciton_IC}
\end{eqnarray}
The $l_{1}$-norm is given by the off-diagonal elements, which read 
\begin{eqnarray}
\left|\rho_{d+}^{e} \right| & = & \frac{1}{2}r, 
\nonumber \\
\left|\rho_{+-}^{e} \right| & = & \frac{1}{2}\sqrt{1-r^{2}}, 
\nonumber \\
\left|\rho_{d-}^{e} \right| & = & \frac{1}{2}r\sqrt{1-r^{2}},
\end{eqnarray}
where we have defined $r=\left|\sin(\vartheta/2)\right|$. We obtain 
the $l_{1}$-norm
\begin{eqnarray}
C_{l_{1}}^{e} = r+\sqrt{1-r^{2}}+r\sqrt{1-r^{2}} \leq \frac{1}{2}+\sqrt{2} \approx 1.9142 
\end{eqnarray}
with equality for $r=\frac{1}{\sqrt{2}}$. This is slightly less than the $l_{1}$-norm
of the maximally coherent state.

REOC is determined by the diagonal elements of $\ket{1} \bra{1}$ in the exciton basis, 
which are 
\begin{eqnarray}
\rho_{dd}^{e}  & = & \frac{1}{2}, 
\nonumber \\
\rho_{++}^{e}  & = & \frac{r^{2}}{2}, 
\nonumber \\
\rho_{--}^{e}  & = & \frac{1-r^{2}}{2}.
\end{eqnarray}
We obtain 
\begin{eqnarray}
C_{{\rm{REOC}}}^{e}  & = & -\frac{1}{2}\log_{2}\frac{1}{2}-\frac{r^{2}}{2}\log_{2}\frac{r^{2}}{2} 
\nonumber \\
 &  & -\frac{1-r^{2}}{2}\log_{2}\frac{1-r^{2}}{2}
\nonumber \\
 & = & 1-\frac{r^{2}}{2}\log_{2}r^{2}-\frac{1-r^{2}}{2}\log_{2}(1-r^{2})
\nonumber \\
 & = & 1+\frac{1}{2}h\left(r^{2}\right)\leq\frac{3}{2}
\end{eqnarray}
with equality again for $r=\frac{1}{\sqrt{2}}$. 

The maximal value is slightly less than that of the maximally coherent trimer state in the 
exciton basis. The condition $r=\frac{1}{\sqrt{2}}$ can be expressed in terms of the original 
Hamiltonian parameters as 
\begin{eqnarray}
\widetilde{E} - \sigma J_{13} = 0, \ J_{23} \neq 0. 
\end{eqnarray}
Thus, given the restriction to perfect population transfer, the coherence in the exciton basis 
can take their maximal value even for nonzero energy barrier, due to constructive interference 
between the two pathways opened up by the nonzero couplings between the three sites. This 
should be compared with the dimer case, where maximal coherence in the exciton basis occurs 
only when the energy barrier vanishes.

\subsection{Efficiency and coherence for optimized coherence}

\paragraph{Site basis}
In the trimer case, the Hamiltonian parameter values that optimize $\overline{C}^s(\vec{h},t)$ 
are roughly the same for $l_{1}$-norm and REOC. When only the absolute value (not the sign) of 
the parameters are considered, there are two sets of parameters. They are shown in tables 
\ref{tab:Trimer_tabell_AC_parametrar1} and \ref{tab:Trimer_tabell_AC_parametrar}. Note that 
$E_{13} \neq 0$ for both sets, which means that perfect population transfer cannot occur for 
these parameter values. 

\begin{table}
\centering
\begin{tabular}{|c|c|c|c|c|}
\hline 
$E_{1}-E_{3}$ & $E_{2}-E_{3}$ & $J_{12}$ & $J_{23}$ & $J_{13}$\tabularnewline
\hline 
\hline 
$0.2$ & $-0.8$ & $-0.5$ & $-0.1$ & $-0.5$\tabularnewline
\hline 
$0.2$ & $-0.8$ & $0.5$ & $0.1$ & $-0.5$\tabularnewline
\hline 
$0.2$ & $-0.8$ & $0.5$ & $-0.1$ & $0.5$\tabularnewline
\hline 
$0.2$ & $-0.8$ & -$0.5$ & $0.1$ & $0.5$\tabularnewline
\hline 
\end{tabular}
\caption{\label{tab:Trimer_tabell_AC_parametrar1} Set 1 of Hamiltonian parameters 
(value and sign) for optimizing $\overline{C}^s(\vec{h},t)$ in a trimer. Note that the 
remaining eight sign combinations do not optimize $\overline{C}^s(\vec{h},t)$.}
\end{table}

\begin{table}
\centering
\begin{tabular}{|c|c|c|c|c|}
\hline 
$E_{1}-E_{3}$ & $E_{2}-E_{3}$ & $J_{12}$ & $J_{23}$ & $J_{13}$\tabularnewline
\hline 
\hline 
$1.0$ & $0.8$ & $-0.5$ & $-0.1$ & $-0.5$\tabularnewline
\hline 
$1.0$ & $0.8$ & $0.5$ & $0.1$ & $-0.5$\tabularnewline
\hline 
$1.0$ & $0.8$ & $0.5$ & $-0.1$ & $0.5$\tabularnewline
\hline 
$1.0$ & $0.8$ & $-0.5$ & $0.1$ & $0.5$\tabularnewline
\hline 
\end{tabular}
\caption{\label{tab:Trimer_tabell_AC_parametrar} Set 2 of Hamiltonian parameters 
(value and sign) for the two sets optimizing $\overline{C}^s(\vec{h},t)$ in a trimer. 
Note that the remaining eight sign combinations do not optimize $\overline{C}^s(\vec{h},t)$.}
\end{table}

\begin{table}
\centering 
\begin{tabular}{|c|c|c|}
\hline 
 & Set $1$ & Set $2$\tabularnewline
\hline 
\hline 
$\overline{C}_{l_{1}}^{12}$ & $0.54$ & $0.66$ \tabularnewline
\hline 
$\overline{C}_{l_{1}}^{23}$ & $0.56$ & $0.56$\tabularnewline
\hline 
$\overline{C}_{l_{1}}^{13}$ & $0.66$ & $0.54$\tabularnewline
\hline 
\end{tabular}
\caption{\label{tab:Trimer_tabell_LAC} Local coherence $\overline{C}_{l_{1}}^{12}(\vec{h},t)$ , 
$\overline{C}_{l_{1}}^{23}(\vec{h},t)$ and $\overline{C}_{l_{1}}^{13}(\vec{h},t)$ for the two sets
of parameters optimizing $\overline{C}^s(\vec{h},t)$ in a trimer.}
\end{table}

\begin{figure}[htb!]
\centering
\includegraphics[width=0.47\textwidth]{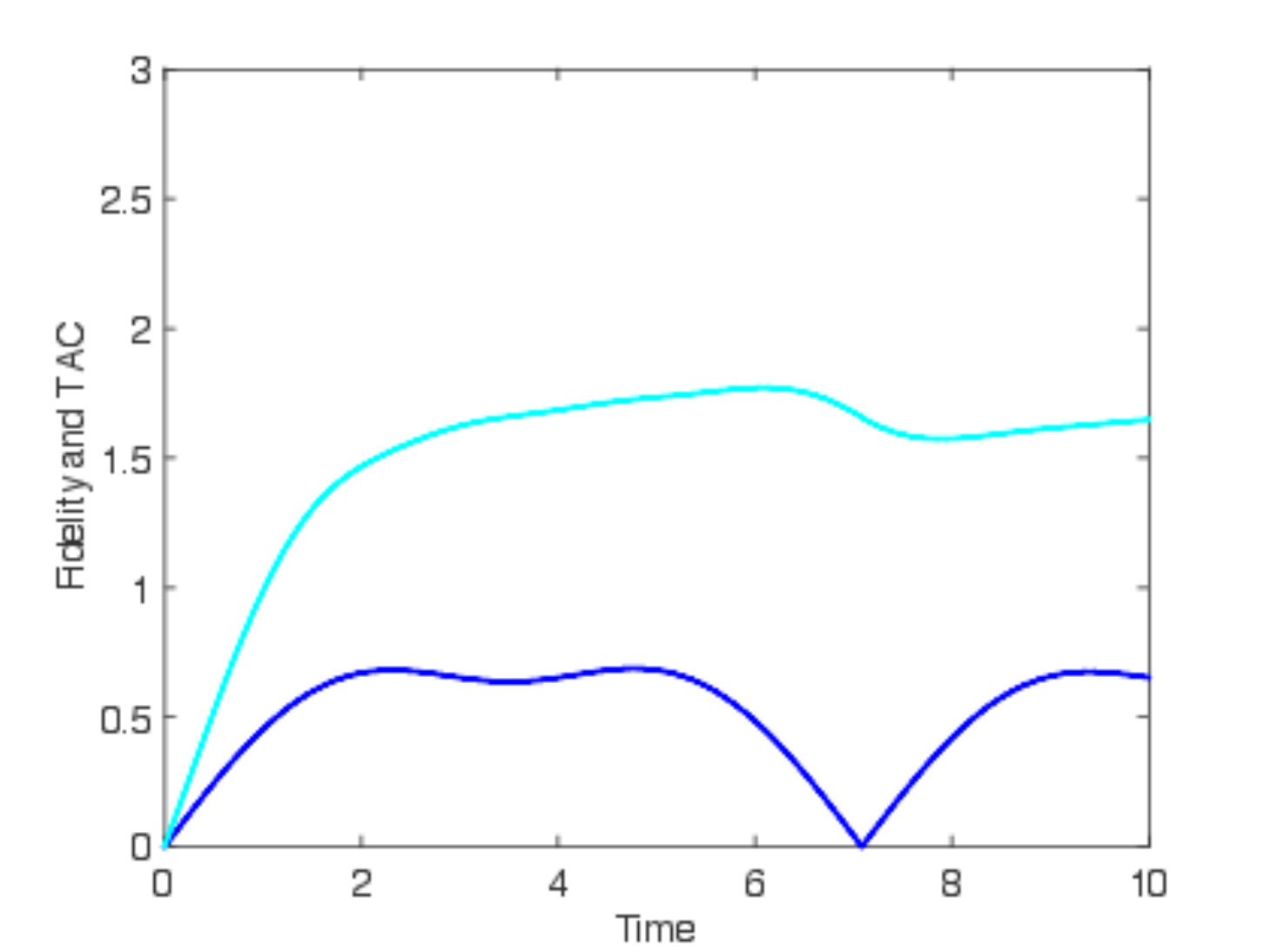}\phantom{.}
\includegraphics[width=0.47\textwidth]{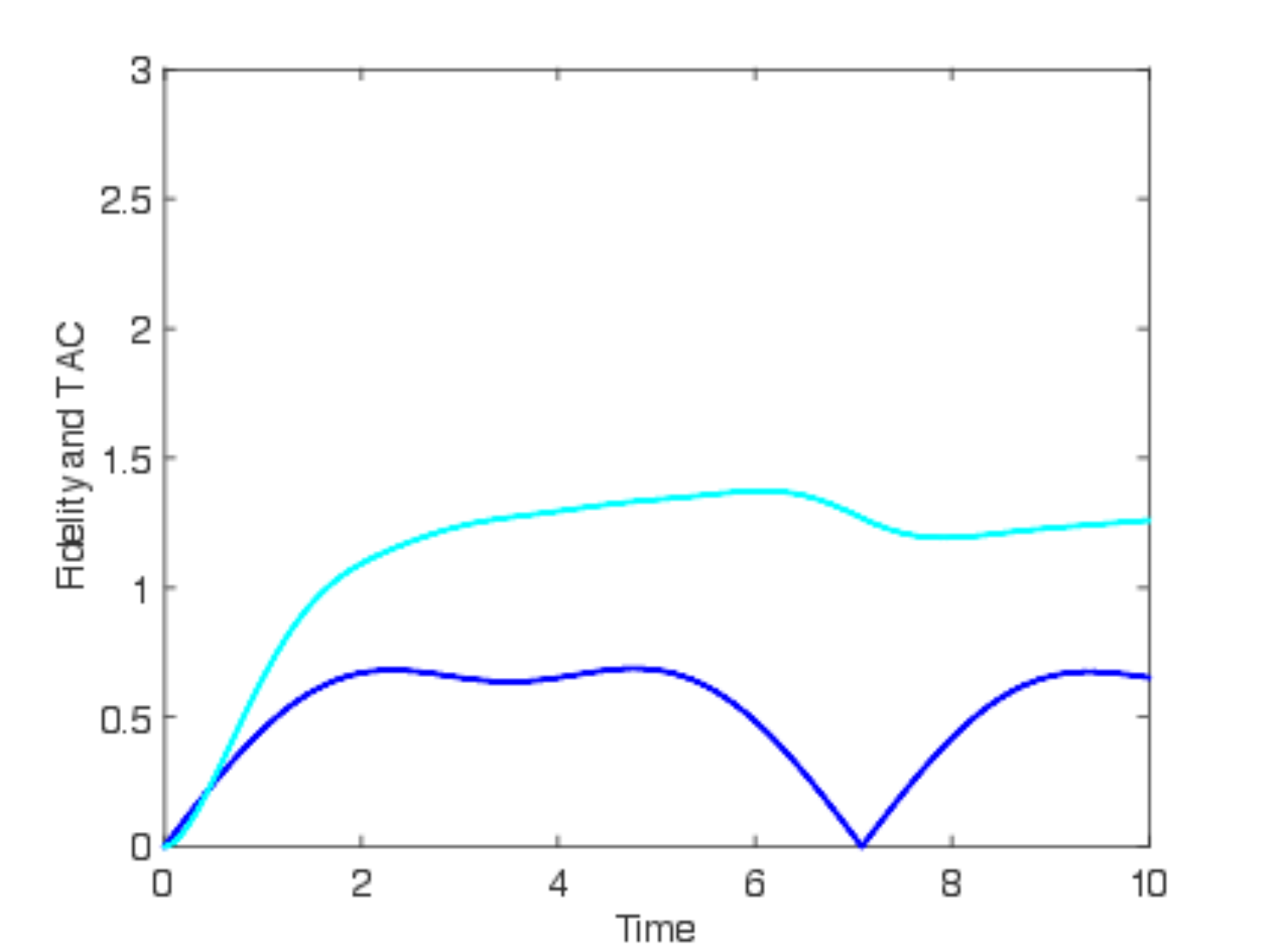}\phantom{.}
\caption{Efficiency and time-averaged coherence for a Hamiltonian
that optimizes  $\overline{C}^s (\vec{h},t)$ in a trimer. The left panel
shows $F(\vec{h},t)$ (blue) and  $\overline{C}_{l_1}^s (\vec{h},t)$
(cyan). The right panel shows $F(\vec{h},t)$ (blue) and $\overline{C}_{\rm{REOC}}^{s} (\vec{h},t)$
(cyan).}
\label{fig:Trimer_site_F_vs_AC_AC_max}
\end{figure}

\begin{figure}[htb!]
\centering
\includegraphics[width=0.47\textwidth]{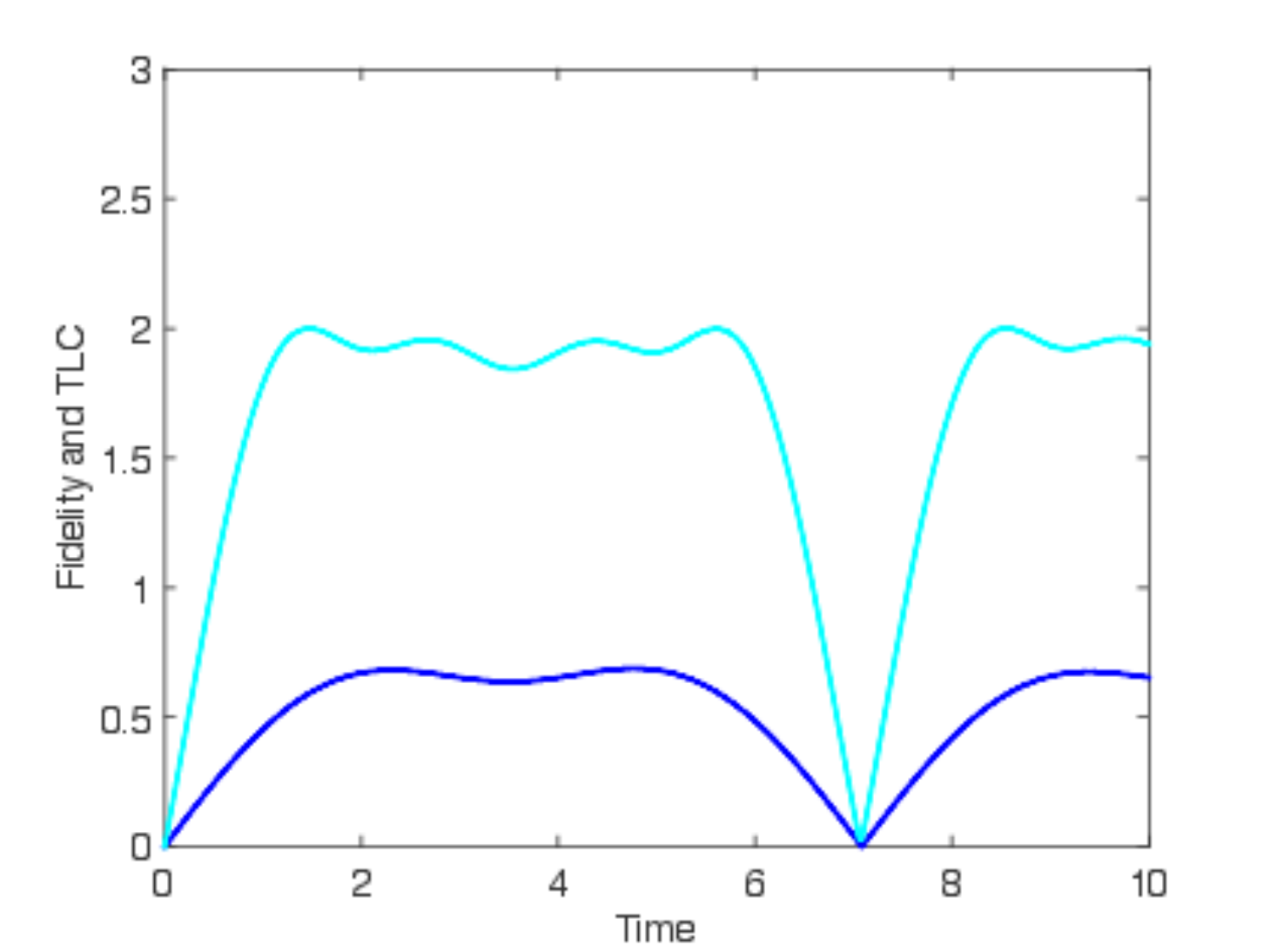}\phantom{.}
\includegraphics[width=0.47\textwidth]{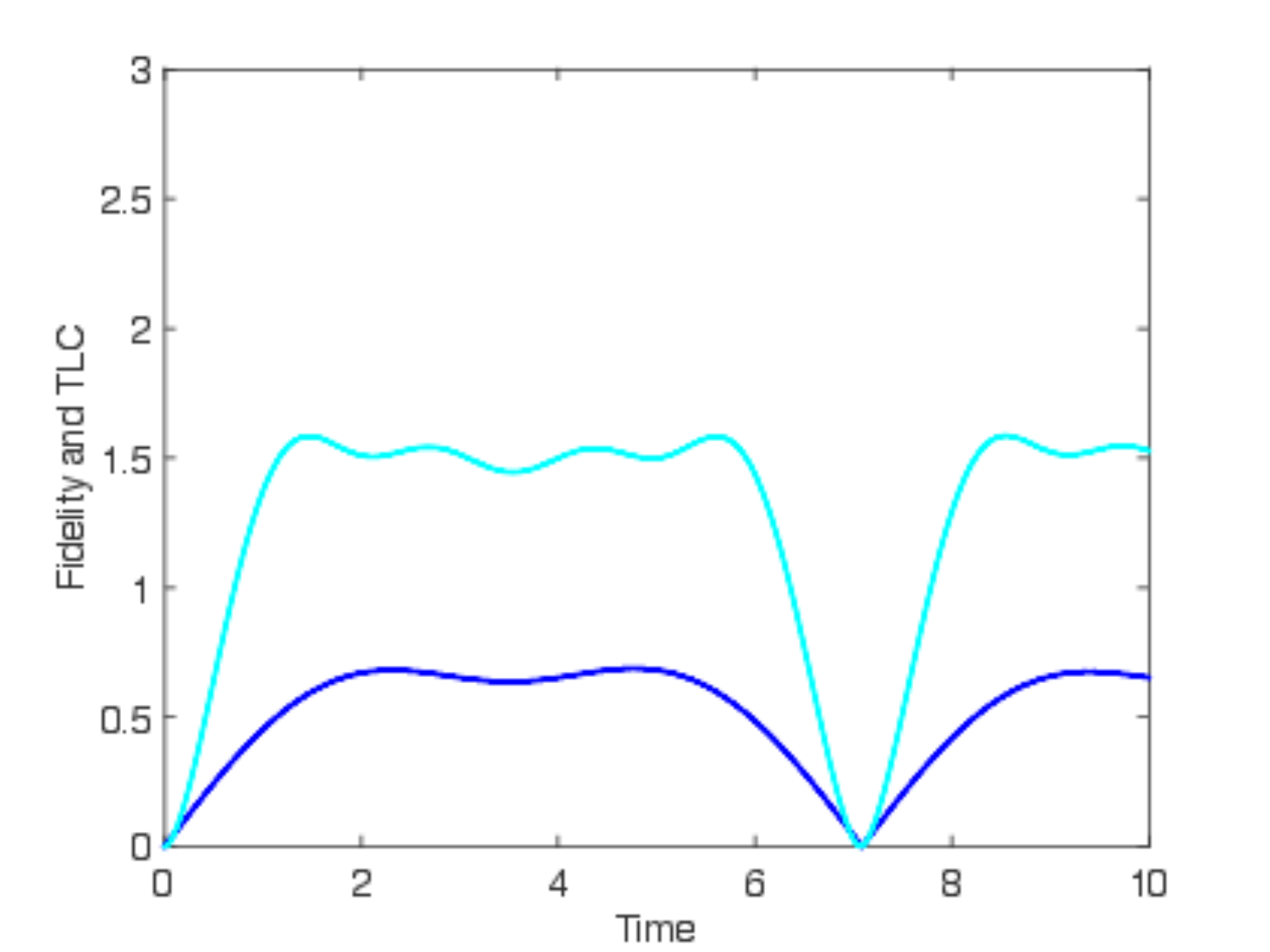}\phantom{.}
\caption{Efficiency and time-local coherence for a Hamiltonian that
optimizes $\overline{C}^s (\vec{h},t)$ in a trimer. The left panel
shows $F(\vec{h},t)$ (blue) and $C_{l_1}^{s} (\vec{h},t)$ (cyan).
The right panel shows $F(\vec{h},t)$ (blue) and $C_{\rm{REOC}}^s (\vec{h},t)$ (cyan).}
\label{fig:Trimer_site_F_vs_C_AC_max}
\end{figure}

\begin{figure}[htb!]
\centering
\includegraphics[width=0.47\textwidth]{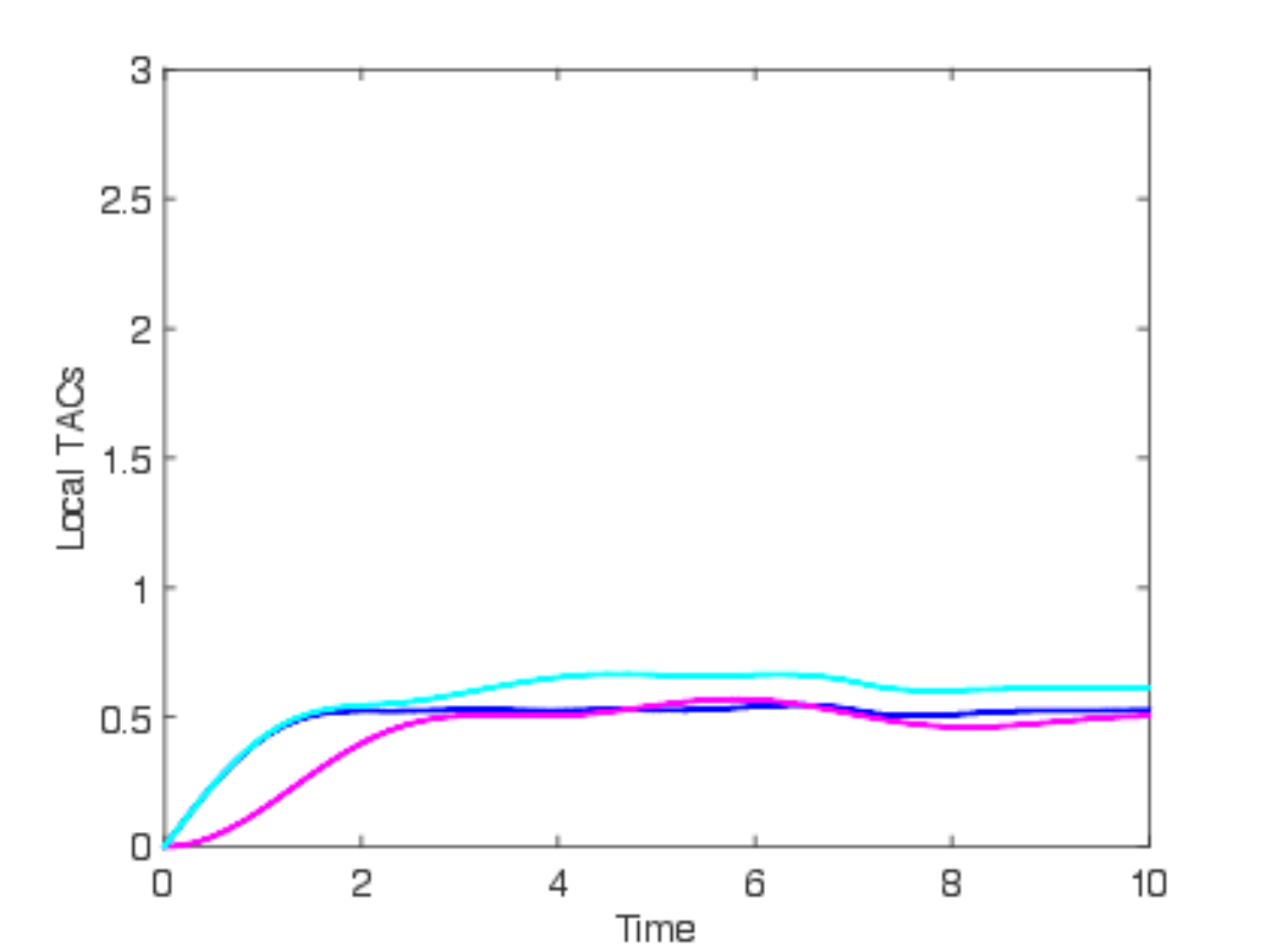}\phantom{.}
\includegraphics[width=0.47\textwidth]{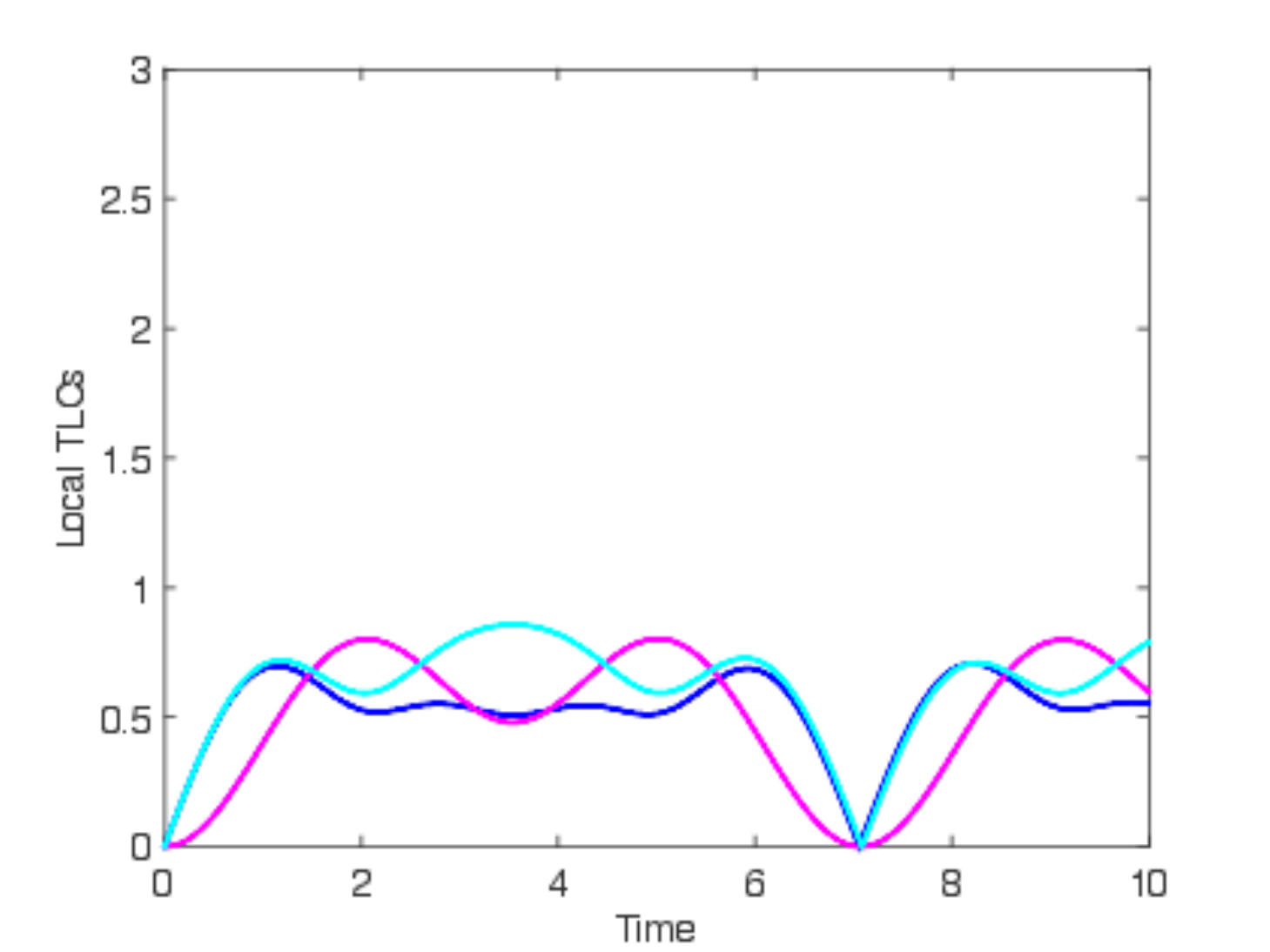}\phantom{.}
\caption{Local coherences  for a Hamiltonian that
optimizes $\overline{C}^s (\vec{h},t)$ in a trimer. The left panel shows $\overline{C}_{l_{1}}^{12} 
(\vec{h},t)$ (blue), $\overline{C}_{l_{1}}^{23} (\vec{h},t)$ (magneta) and 
$\overline{C}_{l_{1}}^{13} (\vec{h},t)$ (cyan).  The right panel shows $C_{l_{1}}^{12} (\vec{h},t)$ (blue), 
$C_{l_{1}}^{23} (\vec{h},t)$ (magneta) and $C_{l_{1}}^{13} (\vec{h},t)$ (cyan).}
\label{fig:Trimer_site_LAC_AC_max}
\end{figure}

The two sets of parameters optimizing $\overline{C}^s(\vec{h},t)$ differ from each other. 
All of them have the same maximal value for $\overline{C}^s (\vec{h},t)$ and $C^s (\vec{h},t)$, 
but $F_{\max}$ is higher in set 1 ($F_{\max} = 0.69$) than in set 2 ($F_{\max} = 0.60$). 
Also, the maximal local time-averaged coherence differs between the two sets, see 
table \ref{tab:Trimer_tabell_LAC}. As can be seen, $\overline{C}_{l_{1}}^{12}(\vec{h},t)$ and
$\overline{C}_{l_{1}}^{13}(\vec{h},t)$ are interchanged between the two sets. 

In figures \ref{fig:Trimer_site_F_vs_AC_AC_max} and \ref{fig:Trimer_site_F_vs_C_AC_max},
$\overline{C}^s(\vec{h},t)$ and $C^s(\vec{h},t)$, respectively, and $F(\vec{h},t)$ for the Hamiltonian 
parameters in set 1 can be seen. Note that all parameter settings in the set give the same 
time-dependence. Again, maximum of $\overline{C}^s (\vec{h},t)$ and maximum of 
$F(\vec{h},t)$ do not coincide in time, but $C^s(\vec{h},t)$ and $F(\vec{h},t)$ are zero 
simultaneously. Note that the upper bounds $C_{l_1}^s = 2$ and 
$C_{{\rm{REOC}}}^s = \log_2 3 \approx 1.58$ are reached in this parameter set.

Local coherence as a function of time for the $l_{1}$-norm (which coincides with concurrence) 
are shown in figure \ref{fig:Trimer_site_LAC_AC_max}. It can be seen that even though for this set, 
$\overline{C}_{l_{1}}^{13} (\vec{h},t)$ reaches a larger value than $\overline{C}_{l_{1}}^{12} 
(\vec{h},t)$ and $\overline{C}_{l_{1}}^{23} (\vec{h},t)$, $C_{l_{1}}^{13} (\vec{h},t)$ and 
$C_{l_{1}}^{12} (\vec{h},t)$ have local minima at the time for maximum of $F(\vec{h},t)$ 
while $C_{l_{1}}^{23} (\vec{h},t)$ has a maximum. Table \ref{tab:Trimer_tabell_LAC} also 
reveals that $\overline{C}_{l_{1}}^{23} (\vec{h},t)$ has the same maximal value in both 
parameter sets. 

\paragraph{Exciton basis}
Optimization of $\overline{C}^e (\vec{h})$ yields infinitely many Hamiltonian parameter sets, 
but none of them coincides with the parameter sets of optimal $F(\vec{h},\tau)$ or optimal 
$\overline{C}^s (\vec{h},t)$. Whether there exists a basis where the $l_{1}$-norm
coincides with fidelity, as in the dimer case, remains an open question. 

\section{Long-term behavior of time-averaged coherence}
\label{sec:long-term}
In section \ref{sec:dimer}, it can be seen in the figures that the time-averaged 
coherence in the site basis seem to converge to a specific limit over long times. 
In this section we discuss this behaviour and extend the time frame of calculations of 
time-averaged coherence.

The evolution of the dimer state is $\pi/\omega$-periodic in $t$ and time-inversion 
symmetric within each period. It thus follows that 
\begin{eqnarray} 
\int_0^{n\frac{\pi}{\omega}} C^s (\hat{\rho} (t')) dt' = 2n\int_0^{\frac{\pi}{2\omega}} 
C^s (\hat{\rho} (t')) dt' ,
\end{eqnarray}
which implies 
\begin{eqnarray}
 & \overline{C}^{s} \left( n\frac{\pi}{\omega}+\delta t ; \hat{\rho} \right) = 
\frac{1}{n\frac{\pi}{\omega} + \delta t} \int_0^{n\frac{\pi}{\omega} + \delta t} C^s (\hat{\rho} (t')) dt'
\nonumber \\ 
 & = \frac{1}{n\frac{\pi}{\omega} + \delta t} \left( 2n \int_0^{\frac{\pi}{2\omega}} 
C^s (\hat{\rho} (t')) dt' + \int_0^{\delta t}C^s (\hat{\rho} (t')) dt' \right) . 
\end{eqnarray}
By assuming $0 \leq \delta t \leq \frac{\pi}{\omega}$, we find 
\begin{eqnarray}
\lim_{n\rightarrow\infty} \overline{C}^{s} \left( n\frac{\pi}{\omega}+\delta t ; \hat{\rho} \right) = 
\frac{2\omega}{\pi} \int_0^{\frac{\pi}{2\omega}} C^s (\hat{\rho} (t')) dt' = 
\overline{C}^{s} \left( \tau; \hat{\rho} \right), 
\end{eqnarray}
which entails that the long-term behavior is determined by the time-averaged coherence 
over one transfer period. It is therefore sufficient to examine the time-averaged $l_1$-norm 
and REOC at $t = \pi /(2\omega) = \tau$.

The time-averaged $l_{1}$-norm measure of coherence over one transfer period in the 
site basis reads
\begin{eqnarray}
\overline{C}_{l_{1}}^{s} (t;\hat{\rho}) \equiv \overline{C}_{l_{1}}^{s} (\theta ,t) = 
\frac{2}{t} \int_0^{t} \sqrt{\cos^2 \theta+
\sin^{2}\theta\cos^2 \omega t'} \sin\theta |\sin \omega t'| dt' .
\end{eqnarray}
We evaluate $\overline{C}_{l_{1}}^{s} (\theta,t)$ at $t = \pi /(2\omega) = \tau$ by making the variable 
substitution $t' \mapsto x = \sin\theta \cos \omega t'$,
yielding
\begin{eqnarray}
 & \overline{C}_{l_{1}}^{s} \left( \theta,\tau \right) = \frac{4}{\pi} 
\int_{0}^{\sin\theta} \sqrt{\cos^{2}\theta+x^{2}} dx 
\nonumber \\
 & = 
\frac{2}{\pi}\left[\sin\theta+\cos^{2}\theta\ln\left(\frac{1+\sin\theta}{\cos\theta}\right)\right] 
\nonumber \\
 & = \frac{2}{\pi} \left[ F_{\max}(\theta)+\left(1-\left[F_{\max}(\theta)\right]^{2}\right) 
\ln\left(\frac{1+F_{\max}(\theta)}{\sqrt{1-\left[ F_{\max}(\theta) \right]^{2}}} \right)\right] .
\nonumber \\
\end{eqnarray}
We thus find 
\begin{eqnarray}
\overline{C}_{l_{1}}^{s} \left( \theta,\tau \right) = \frac{2}{\pi} 
\end{eqnarray}
in the case of perfect population transfer $F_{\max} = 1$, and
\begin{eqnarray}
\overline{C}_{l_{1}}^{s} \left( \theta,\tau \right) \approx \frac{2}{\pi} 1.12 
\end{eqnarray}
corresponding to $F_{\max}(\tau) \approx 0.83$ (the slight difference between this value and the 
one corresponding to table \ref{tab:Dimer_tabell} shows that the asymptotic value differs from 
the optimal value of the time-averaged coherence), when $\overline{C}_{l_{1}}^{s}(\theta,\tau)$ 
is maximal. 

Similarly, the time-averaged REOC for perfect population transfer tends asymptotically to 
\begin{eqnarray}
\overline{C}_{\rm{REOC}}^{s} \left( \theta,\tau \right) = 
\frac{2\omega}{\pi} \int_0^{\frac{\pi}{2\omega}} h\left( \left[ F(\theta,t') \right]^2 \right) dt' . 
\end{eqnarray}
Since there is no simple analytic solution to this integral, we resort only to numerical 
solutions in this case. 

\begin{figure}[htb!]
\centering
\includegraphics[width=0.47\textwidth]{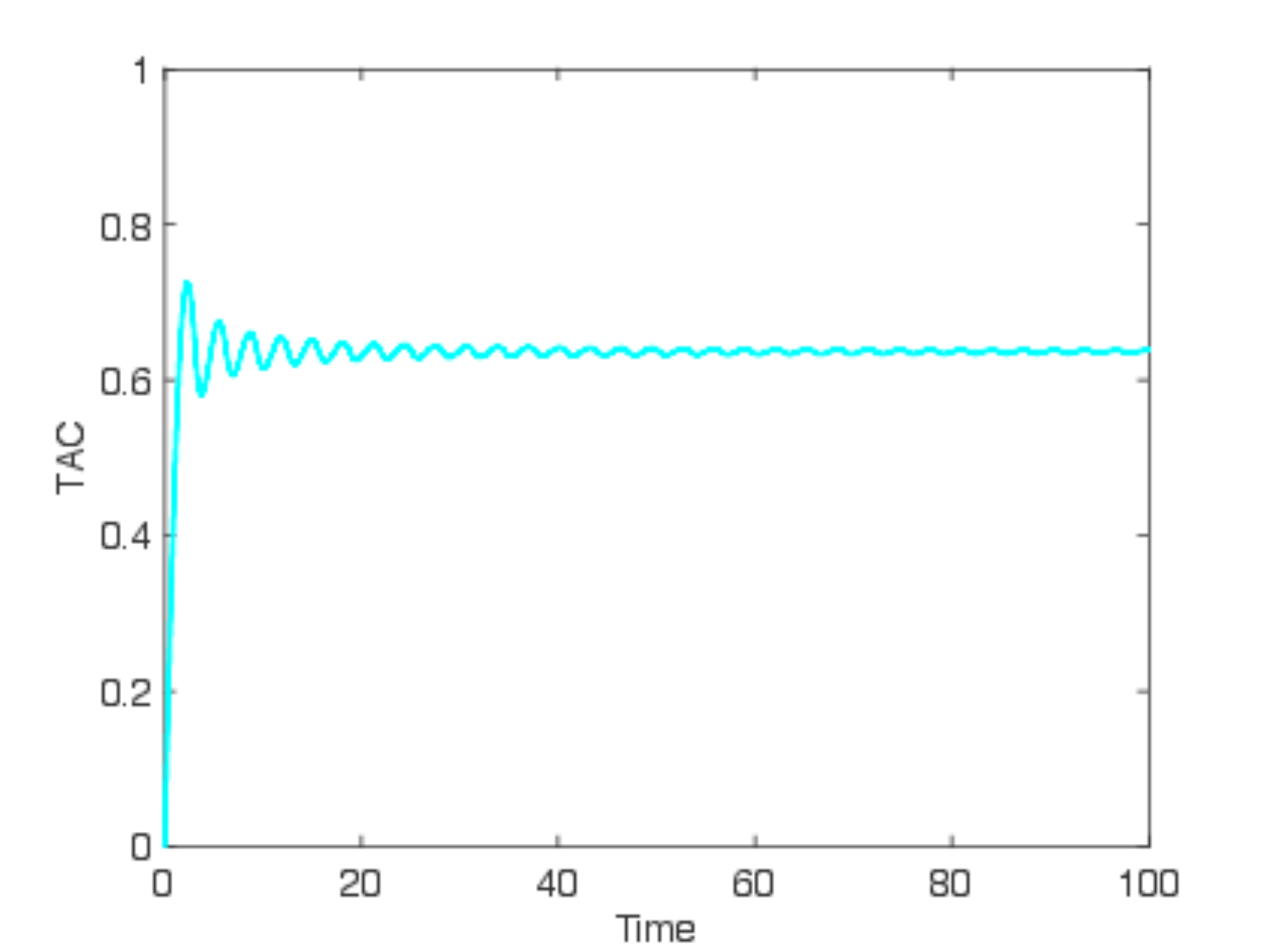}\phantom{.}
\includegraphics[width=0.47\textwidth]{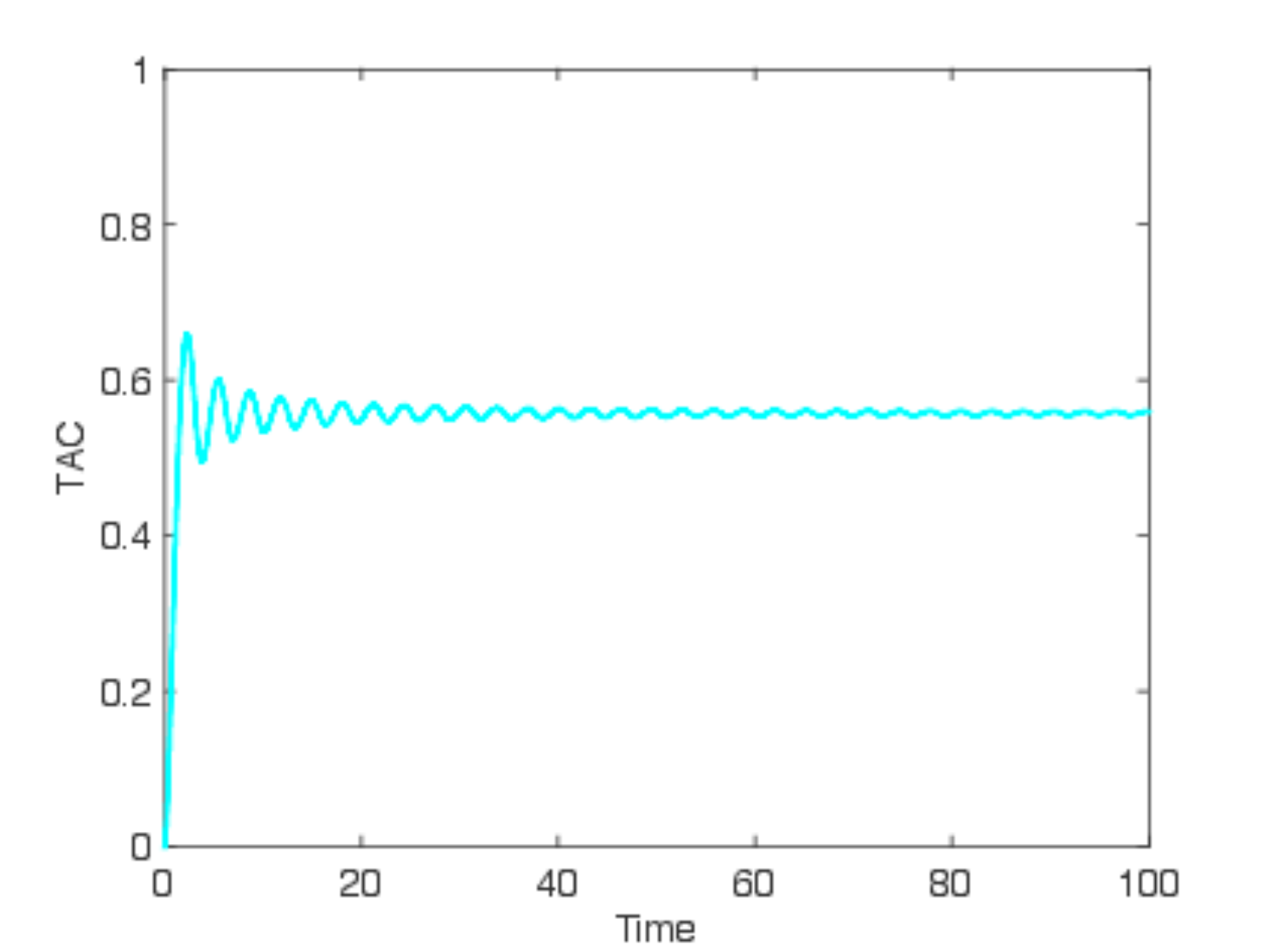}\phantom{.}
\caption{Long time behaviour of time-averaged coherence for a Hamiltonian
that optimizes $F(\theta,\tau)$ in a dimer. The left panel shows $\overline{C}_{l_1}^s(\theta,t)$
and the right panel shows $\overline{C}_{\rm{REOC}}^s(\theta,t)$ .}
\label{fig:Dimer_site_medelkoherens_F_max}
\end{figure}

\begin{figure}[htb!]
\centering
\includegraphics[width=0.47\textwidth]{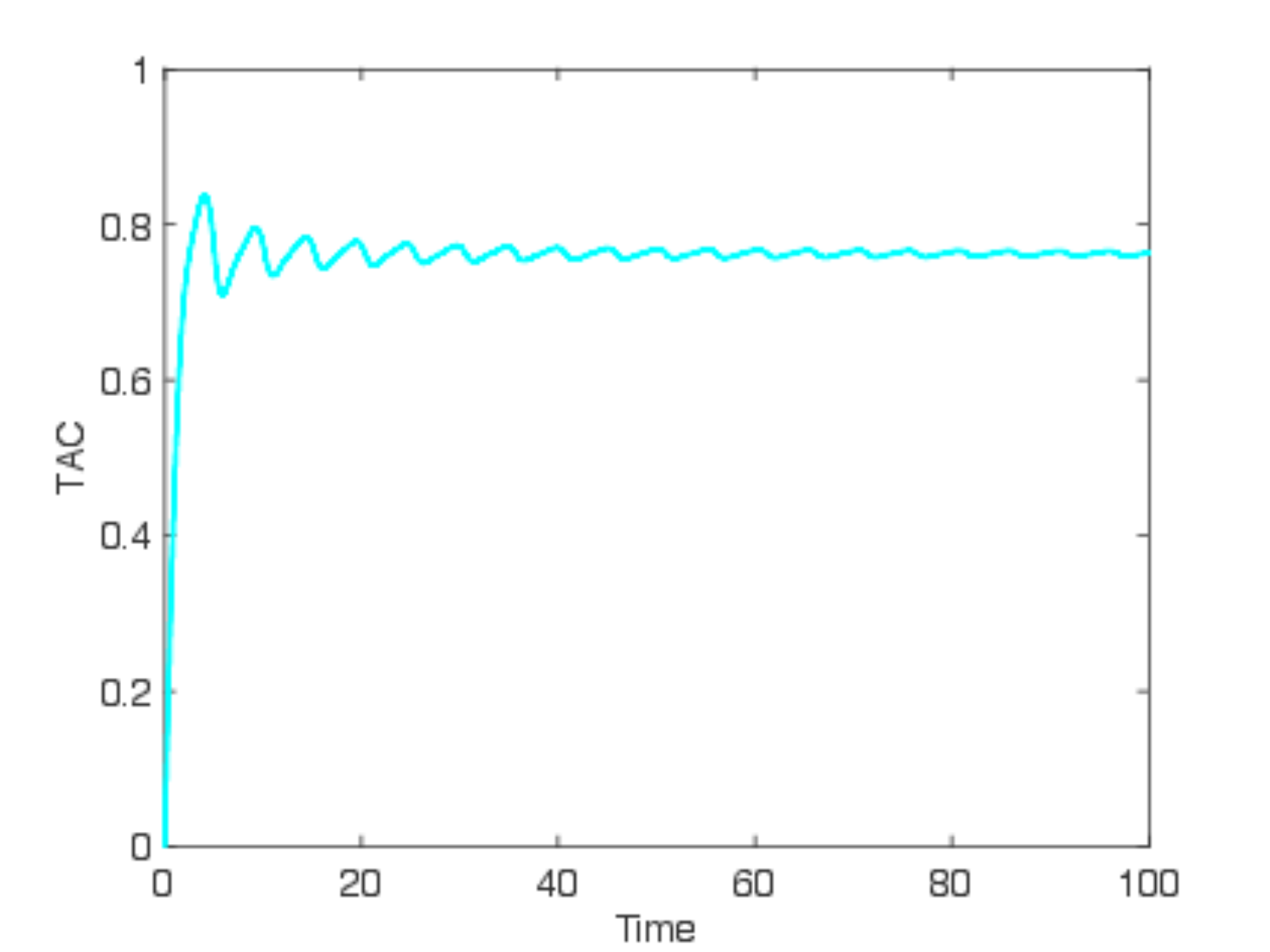}\phantom{.}
\includegraphics[width=0.47\textwidth]{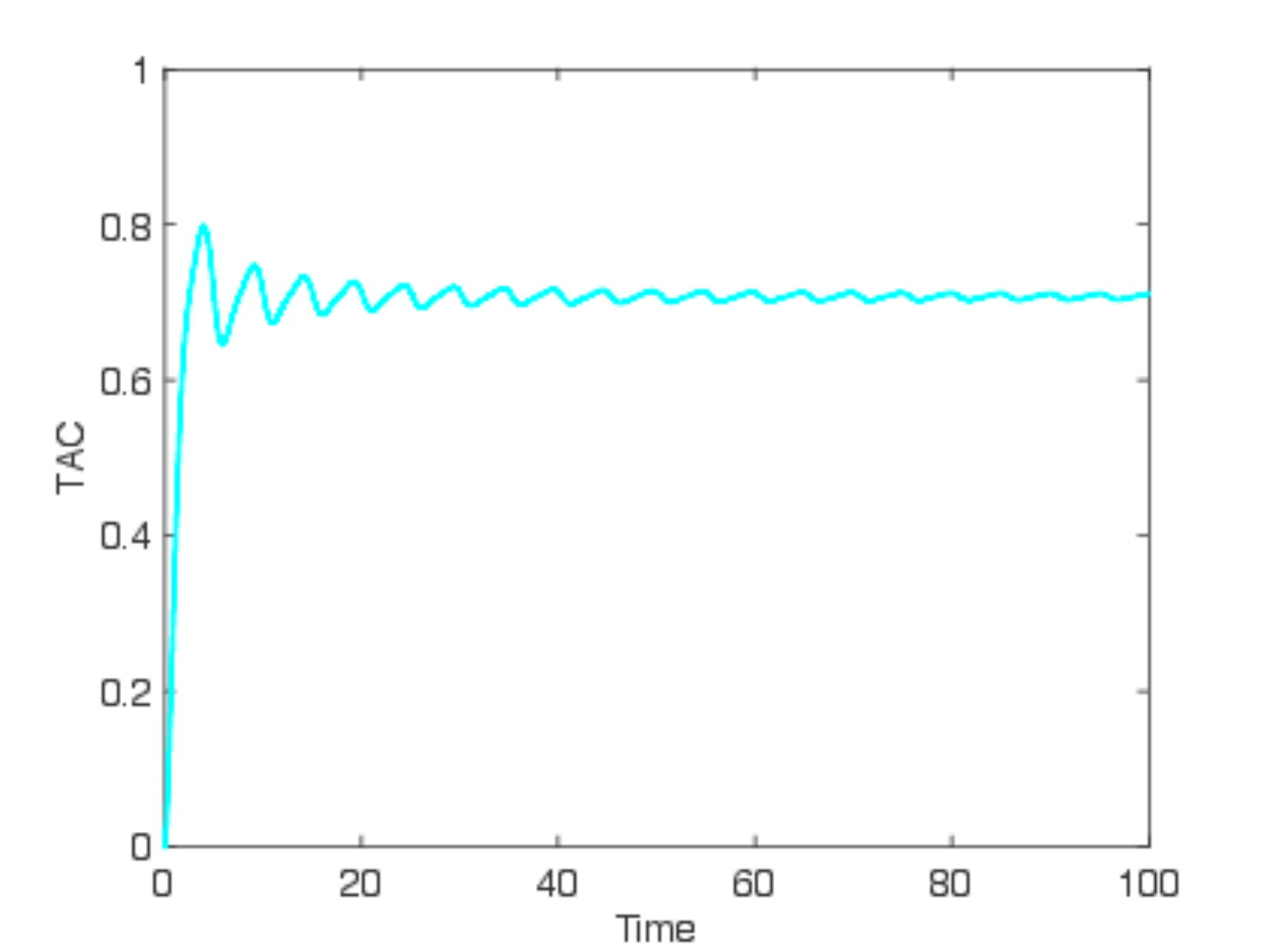}\phantom{.}
\caption{Long time behaviour of time-averaged coherence for a Hamiltonian
that optimizes $\overline{C}^s(\theta,t)$  in a dimer. The left panel
shows $\overline{C}_{l_1}^s(\theta,t)$ and the right panel shows 
$\overline{C}_{\rm{REOC}}^s(\theta,t)$.}
\label{fig:Dimer_site_medelkoherens_AC_max}
\end{figure}

Simulation of the long-term behaviour of $\overline{C}_{l_{1}}^s(\theta,t)$ and 
$\overline{C}_{{\rm{REOC}}}^s(\theta,t)$ for the parameter values that optimize 
$F(\theta,\tau)$ and $\overline{C}^s(\theta,t)$ are shown in figures 
\ref{fig:Dimer_site_medelkoherens_F_max} and \ref{fig:Dimer_site_medelkoherens_AC_max}, 
respectively. The trend towards an asymptotic value is clearly visible in each case. 

\begin{figure}[htb!]
\centering
\includegraphics[width=0.47\textwidth]{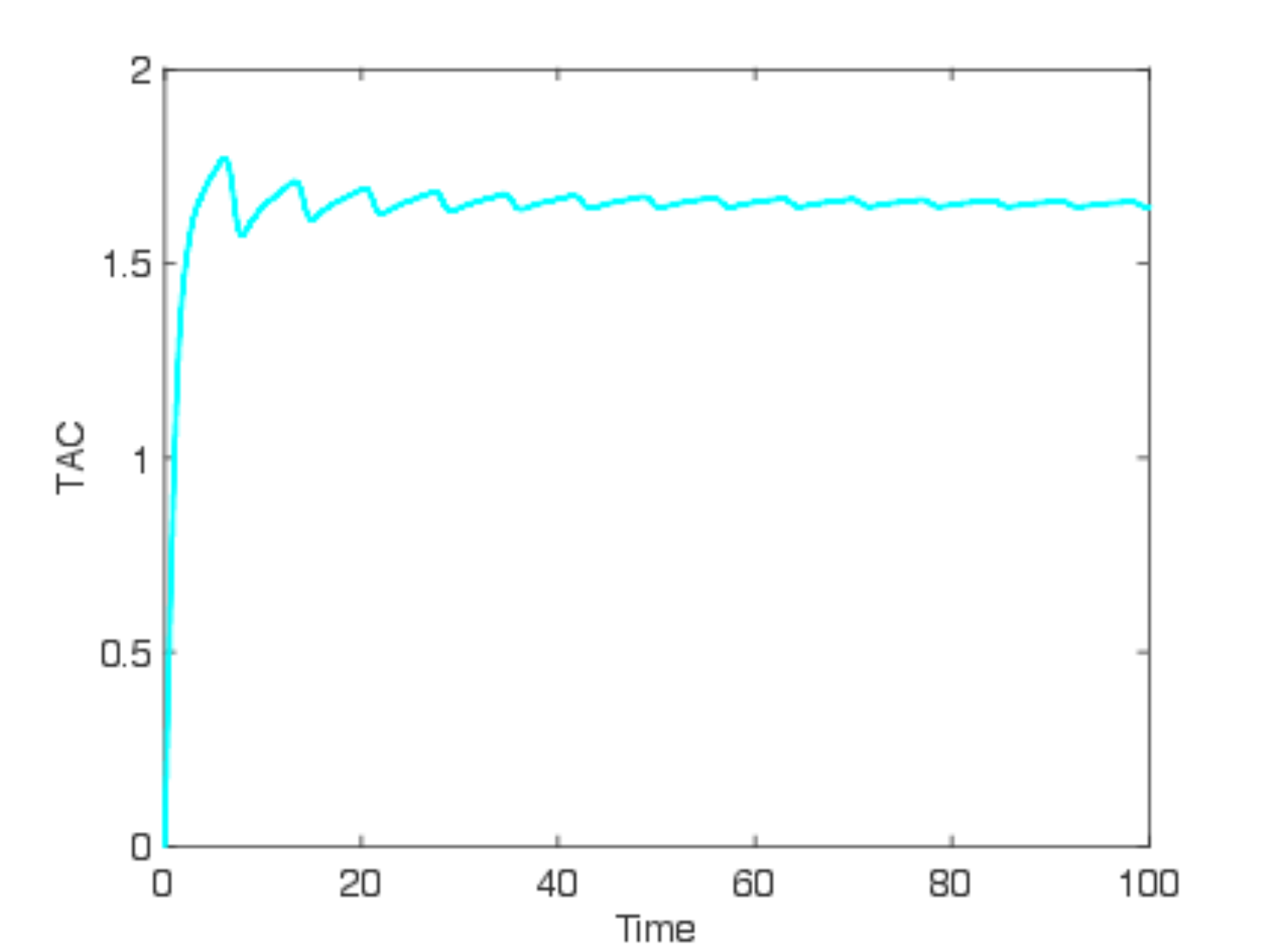}\phantom{.}
\includegraphics[width=0.47\textwidth]{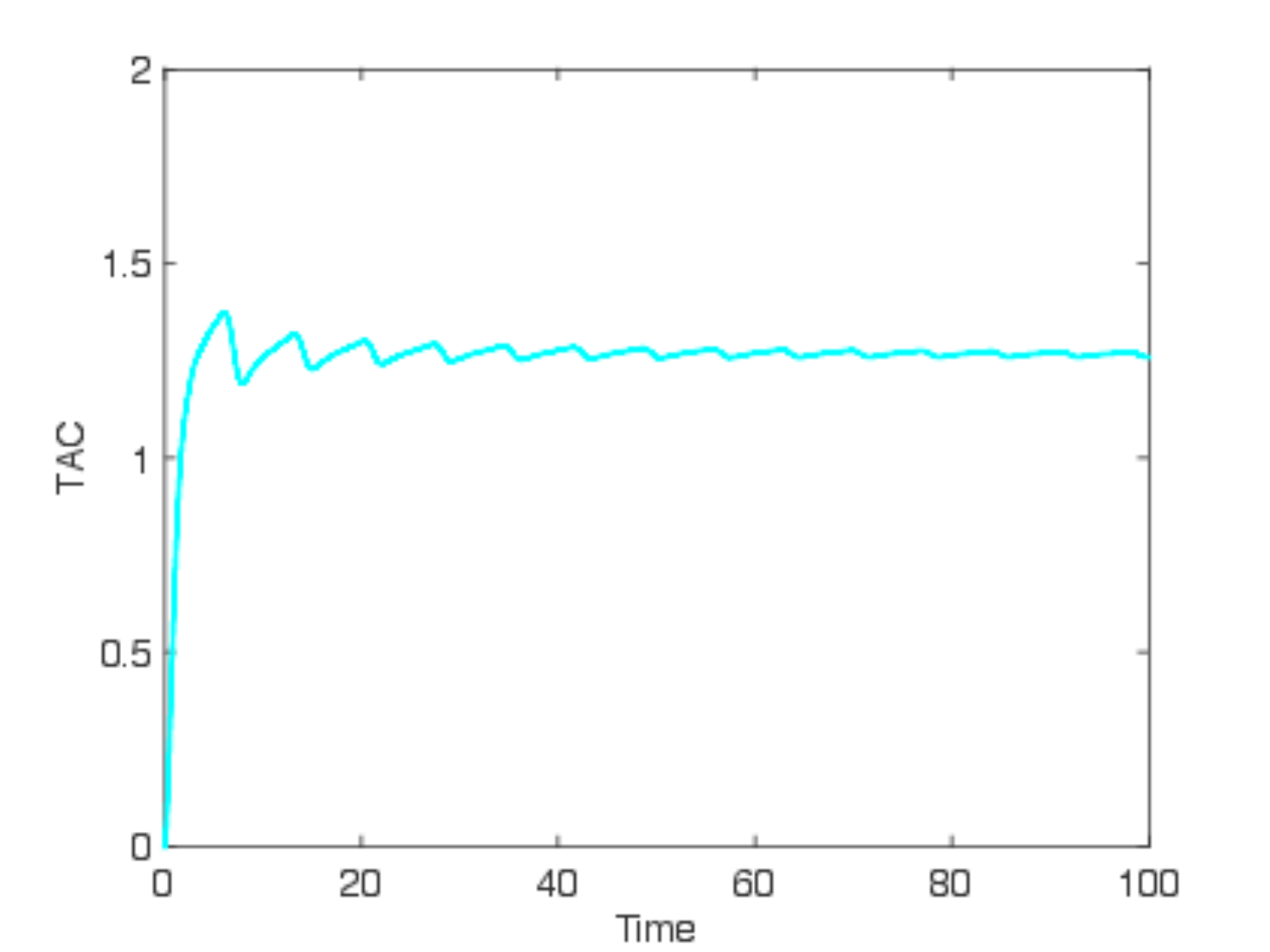}\phantom{.}
\caption{Long time behaviour of time-averaged coherence for a Hamiltonian
that optimizes $\overline{C}^s (\vec{h},t)$ in a trimer. The left
panel shows  $\overline{C}_{l_1}^s(\vec{h},t)$ and the right panel
shows $\overline{C}_{\rm{REOC}}^s(\vec{h},t)$.}
\label{fig:Trimer_site_medelkoherens_AC_max}
\end{figure}

The long-term behaviour of $\overline{C}_{l_{1}}^s(\vec{h},t)$ and $\overline{C}_{{\rm {REOC}}}^s 
(\vec{h},t)$ for parameter values optimizing $\overline{C}^s(\vec{h},t)$ in the trimer is shown in 
figure \ref{fig:Trimer_site_medelkoherens_AC_max}. We note a trend towards an asymptotic 
value also in the trimer case.  

\section{Conclusions}
In this study we have numerically investigated the relation between efficient population transfer 
and coherence - in different bases and quantified by different coherence measures - in a dimer 
and trimer system undergoing Schr\"odinger evolution. Optimization has been performed 
over a parameter space defined by the site energies and inter-site couplings of a tight-binding 
Hamiltonian. A possible realization of such a system could be a molecular aggregate of 
chromophores occupying the sites and being described as two-level systems. 

We have found several relations between the efficiency and coherence measures used to study 
the population transfer process. We summarize these findings as follows: 

\begin{itemize}

\item The maximum of the time-local coherence and the population transfer in the dimer 
case occur simultaneously for a maximal transfer efficiency below a certain threshold 
value. Above this value the coherence maximum precedes the fidelity maximum. For perfect 
population transfer, the coherence becomes maximal at precisely halfway before and 
vanishes at the completion of the transfer. Thus, efficient population transfer is characterized 
by a time-local coherence fully localized before the transfer has been completed. 

\item It has been shown that in neither the dimer nor the trimer do the parameter values 
of maximal time-averaged coherence in the site basis coincide with the parameter values 
corresponding to perfect population transfer between the end-sites. In other words, the efficiency 
and time-averaged coherence order the Hamiltonians differently with respect to their ability to 
transfer population. 

\item In the dimer, the population transfer efficiency is in one-to-one correspondence 
with coherence in the exciton basis. Thus, perfect population transfer coincides with the 
upper bounds of the coherence. In this sense, coherence in the exciton basis is directly 
linked to the efficiency of the population transfer. In the trimer system, the time-averaged 
coherence in the exciton basis is about $95\%$ of its upper bound at perfect population 
transfer; thus, the one-to-one correspondence between efficiency and coherence seems 
to be restricted to the dimer case. 

\item Optimal efficiency and time-averaged coherence in the site basis do not in general coincide 
in time. Hence, maximal efficiency and maximal time-averaged coherence are not obtained 
simultaneously in a system without environmental interactions. This result indicates that 
coherence in the site basis on its own plays no immediate role for efficient population transfer. 

\end{itemize}

The present analysis can be extended to more sites and environmental effects. This would 
make it possible to examine the potential impact of quantum coherence on the efficiency of 
population transport in various systems, such as EET in photosynthetic complexes, under 
realistic systems. The possibility of a multitude of pathways in such systems would lead to 
a rich interplay between tunneling and interference effects that can be analyzed and understood 
by quantifying the coherence during the time evolution. 

The related problem of transferring quantum states in spin networks to communicate 
quantum information between quantum registers has attracted considerable 
attention in the past \cite{Bose2007,Kay2010}. It would be of interest to examine the role 
of coherence in such processes, in particular to study the optimization of coherence in 
relation to the state transfer fidelity. 

\section*{Acknowledgments}
The computations were performed on resources provided by the Swedish National 
Infrastructure for Computing (SNIC) at Uppsala Multidisciplinary Center for Advanced 
Computational Science (UPPMAX) under Project snic2017-7-17. E.S. acknowledges 
financial support from the Swedish Research Council (VR) through Grant No. D0413201.

\end{document}